\documentclass[a4paper,UKenglish]{dagrep-v2018}
\usepackage[utf8]{inputenc}
\usepackage{microtype}

\usepackage[utf8]{inputenc}
\usepackage{latexsym}
\usepackage{graphicx}
\graphicspath{{./images/}}
\usepackage{booktabs} 
\usepackage{color}  
\usepackage{amsmath}  
\usepackage{caption}
\usepackage{tikz}
\usepackage{colortbl} 
\usepackage{framed}
\usepackage{multirow}
\usepackage{multicol}
\usepackage{hyperref}
\usepackage{url}
\usepackage{balance}
\usepackage{verbatim}
\usepackage{cancel}
\usepackage{xspace} 
\usepackage[ruled,vlined]{algorithm2e}
\usepackage{bbold} 

\newenvironment{Snugshade}[1][236,236,236]{
    \setlength{\itemsep}{0pt}
     \setlength{\parsep}{0pt}
     \setlength{\topsep}{0pt}
     \setlength{\partopsep}{0pt}
     \setlength{\leftmargin}{1.5em}
     \setlength{\labelwidth}{0em}
     \setlength{\labelsep}{0em} 
\setlength{\parskip}{0pt}
    \definecolor{shadecolor}{RGB}{#1}%
    \begin{snugshade}
}{%
    \end{snugshade}%
}

\newcommand{\squishlist}{
 \begin{list}{$\bullet$}
  { \setlength{\itemsep}{0pt}
     \setlength{\parsep}{1pt}
     \setlength{\topsep}{1pt}
     \setlength{\partopsep}{0pt}
     \setlength{\leftmargin}{1.5em}
     \setlength{\labelwidth}{1em}
     \setlength{\labelsep}{0.5em} } }

\newcommand{\squishend}{
  \end{list}  }

\subject{Report from Dagstuhl Seminar 19461}
\title{Conversational Search}
\titlerunning{19461 -- Conversational Search}

\author[1]{Avishek Anand}
\author[2]{Lawrence Cavedon}
\author[3]{Matthias Hagen}
\author[4]{Hideo Joho}
\author[5]{Mark Sanderson}
\author[6]{Benno Stein}
\authorrunning{Avishek Anand, Lawrence Cavedon, Hideo Joho, Mark Sanderson, and Benno Stein}
\affil[1]{Leibniz Universität Hannover, DE, \texttt{anand@kbs.uni-hannover.de}}
\affil[2]{RMIT University - Melbourne, AU, \texttt{lawrence.cavedon@rmit.edu.au}}
\affil[3]{MLU Halle, DE, \texttt{matthias.hagen@informatik.uni-halle.de}}
\affil[4]{University of Tsukuba - Ibaraki, JP, \texttt{hideo@slis.tsukuba.ac.jp}}
\affil[5]{RMIT University - Melbourne, AU, \texttt{mark.sanderson@rmit.edu.au}}
\affil[6]{Bauhaus-Universität Weimar, DE, \texttt{benno.stein@uni-weimar.de}}

\seminarnumber{19461}
\semdata{November 10-15, 2019 -- \href{https://www.dagstuhl.de/19461}{http://www.dagstuhl.de/19461}}
\subjclass{Artificial Intelligence/Robotics, Data Bases/Information Retrieval, Society/Human-computer Interaction}
\additionaleditors{Al-Khatib, Khalid, Leonhardt, Jurek, Trippas, Johanne}

\volumeinfo
  {Avishek Anand, Lawrence Cavedon, Hideo Joho, Mark Sanderson, and Benno Stein}
  {5}
  {Conversational Search}
  {9}
  {11}
  {1}
\DOI{10.4230/DagRep.9.11.1}

\makeatletter
\renewcommand\license{\@preambleentry{License}{%
  \href{http://creativecommons.org/licenses/by/3.0/}%
       {\raise-2\p@\hb@xt@1.44em{
          \includegraphics[height=7.5\p@,clip]{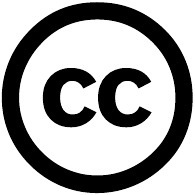}\enskip}}%
  \href{http://creativecommons.org/licenses/by/3.0/}
               {Creative Commons BY 3.0 Unported} license\newline
  \null\hskip0.055em\textsf{\textcopyright}\enskip\nameref{absaut\theabsaut}}}
\makeatother

\makeatletter
\@ifpackagelater{graphicx}{2015/01/01}{
\setkeys{Gin}{pagebox=artbox}}{\pdfpagebox 5}
\makeatother
\DeclareGraphicsExtensions{.pdf,.ai,.jpg,.png}
\graphicspath{{.}}

\begin{document}

\maketitle

\begin{abstract}
Dagstuhl Seminar 19461 ``Conversational Search'' was held on 10-15 November 2019. 44~researchers in Information Retrieval and Web Search, Natural Language Processing, Human Computer Interaction, and Dialogue Systems were invited to share the latest development in the area of Conversational Search and discuss its research agenda and future directions. A 5-day program of the seminar consisted of six introductory and background sessions, three visionary talk sessions, one industry talk session, and seven working groups and reporting sessions. The seminar also had three social events during the program. This report provides the executive summary, overview of invited talks, and findings from the seven working groups which cover the definition, evaluation, modelling, explanation, scenarios, applications, and prototype of Conversational Search. The ideas and findings presented in this report should serve as one of the main sources for diverse research programs on Conversational Search.
\end{abstract}

\newpage
\section{Executive Summary}

\summaryauthor[Avishek Anand, Lawrence Cavedon, Hideo Joho, Mark Sanderson, Benno Stein]{Avishek Anand (Leibniz Universit\"{a}t Hannover, DE)\\ Lawrence Cavedon (RMIT University - Melbourne, AU)\\ Hideo Joho (University of Tsukuba - Ibaraki, JP)\\ Mark Sanderson (RMIT University - Melbourne, AU)\\ Benno Stein (Bauhaus-Universit\"{a}t Weimar, DE)}
\license

\subsection*{Background and Motivation}

The Conversational Search Paradigm promises to satisfy information needs using human-like dialogs, be it in spoken or in written form. This kind of ``information-providing dialogs'' will increasingly happen enpassant and spontaneously, probably triggered by smart objects with which we are surrounded such as intelligent assistants such as Amazon Alexa, Apple Siri, Google Assistant, and Microsoft Cortana, domestic appliances, environmental control devices, toys, or autonomous robots and vehicles. The outlined development marks a paradigm shift for information technology, and the key question(s) is (are):

\textit{What does Conversational Search mean and how to make the most of it---given the possibilities and the restrictions that come along with this paradigm?}

Currently, our understanding is still too limited to exploit the Conversational Search Paradigm for effectively satisfying the existing diversity of information needs. Hence, with this first Dagstuhl Seminar on Conversational Search we intend to bring together leading researchers from relevant communities to understand and to analyze this promising retrieval paradigm and its future from different angles.

Among others, we expect to discuss issues related to interactivity, result presentation, clarification, user models, and evaluation, but also search behavior that can lead into a human-machine debate or an argumentation related to the information need in question.

Moreover, we expect to define, shape, and formalize a set of corresponding problems to be addressed, as well as to highlight associated challenges that are expected to come in the form of multiple modalities and multiple users. Correspondingly, we intend to define a roadmap for establishing a new interdisciplinary research community around Conversational Search, for which the seminar will serve as a prominent scientific event, with hopefully many future events to come.

\subsection*{Seminar Program}

A 5-day program of the seminar consisted of six introductory and background sessions, three visionary talk sessions, one industry talk session, and nine breakout discussion and reporting sessions. The seminar also had three social events during the program. The detail program of the seminar is available online.%
\footnote{\url{https://www.dagstuhl.de/schedules/19461.pdf}}

\subsubsection*{Pre-Seminar Activities}

Prior to the seminar, participants were asked to provide inputs to the following questions and request:
\begin{enumerate}
\item
What are your ideas of the ``ultimate'' conversational search system?
\item
Please list, from the perspective of your research field, important open questions or challenges in conversational search.
\item
What are the three papers a PhD student in conversational search should read and why?
\end{enumerate}

From the survey, the following topics were initially emerged as interests of participants. Many of these topics were discussed at length in the seminar.
\begin{itemize}
\item
Understanding nature of information seeking in the context of conversational agents
\item
Modelling problems in conversational search 
\item
Clarification and explanation
\item
Evaluation in conversational search systems
\item
Ethics and privacy in conversational systems
\item
Extending the problem space beyond the search interface and Q/A
\end{itemize}

Another outcome of the above pre-seminar questions was a compilation of recommended reading list to gain a solid understanding of topics and technologies that were related to the research on Conversational Search. The reading list is provided in Section~\ref{reading-list} of this report.

\subsubsection*{Invited Talks}

One of the main goals and challenges of this seminar was to bring a broad range of researchers together to discuss Conversational Search, which required to establish common terminologies among participants. Therefore, we had a series of 18~invited talk throughout the seminar program to facilitate the understanding and discussion of conversational search and its potential enabling technologies. The main part of this report includes the abstract of all talks.

\subsubsection*{Working Groups}

In the afternoon of Day~2, initial working groups were formed based on the inputs to the pre-seminar questionnaires, introductory and background talks, and discussions among participants. On Day~3, the grouping was revisited and updated, and, eventually, the following seven groups were formed to focus on topics such as the definition, evaluation, modelling, explanation, scenarios, applications, and prototype of Conversational Search.
\begin{itemize}
\item
Defining Conversational Search
\item
Evaluating Conversational Search
\item
Modeling in Conversational Search
\item
Argumentation and Explanation
\item
Scenarios that Invite Conversational Search
\item
Conversation Search for Learning Technologies
\item
Common Conversational Community Prototype: Scholarly Conversational Assistant
\end{itemize}

We have summarized the working groups' outcomes in the following. Please refer to the main part of this report for the full description of the findings.

\paragraph*{Defining Conversational Search}

This group aimed to bring structure and common terminology to the different aspects of conversational search systems that characterise the field. After reviewing existing concepts such as Conversational Answer Retrieval and Conversational Information Seeking, the group offers a typology of Conversational Search systems via functional extensions of information retrieval systems, chatbots, and dialogue systems. The group further elaborates the attributes of Conversational Search by discussing its dimensions and desirable additional properties. Their report suggests types of systems that should not be confused as conversational search systems.

\paragraph*{Evaluating Conversational Search}

This group addressed how to determine the quality of conversational search for evaluation. They first describe the complexity of conversation between search systems and users, followed by a discussion of the motivation and broader tasks as the context of conversational search that can inform the design of conversational search evaluation. The group also surveys 12~recent tasks and datasets that can be exploited for evaluation of conversational search. Their report presents several dimensions in the evaluation such as User, Retrieval, and Dialog, and suggests that the dimensions might have an overlap with those of Interactive Information Retrieval.

\paragraph*{Modeling Conversational Search}

This group addressed what should be modeled from the real world to achieve a successful conversational search and how. They explain why a range of concepts and variables such as capabilities and resources of systems, beliefs and goals of users, history and current status of process, and search topics and tasks should be considered to advance understanding between systems and users in the context of Conversational Search. The group points out that the options the current search engines present to users can be too broad in conversational interaction. They suggest that a deeper modeling of users' beliefs and wants, development of reflective mechanisms, and finding a good balance between macroscopic and microscopic modeling are promising directions for future research.

\paragraph*{Argumentation and Explanation}

Motivated by inevitable influences made to users due to the course of actions and choices of search engines, this group explored how the research on argumentation and explanation can mitigate some of potential biases generated during conversational search processes, and facilitate users' decision-making by acknowledging different viewpoints of a topic. The group suggests a research scheme that consists of three layers: a conversational layer, a demographics layer, and a topic layer. Also, their report explains that argumentation and explanation should be carefully considered when search systems (1)~select, (2)~arrange, and (3)~phrase the information presented to the users. Creating an annotated corpus with these elements is the next step in this direction.

\paragraph*{Scenarios for Conversational Search}

This group aimed to identify scenarios that invite conversational search, given that natural language conversation might not always be the best way to search in some context. Their report summarises that modality and task of search are the two cases where conversational search might make sense. Modality can be determined by a situation such as driving or cooking, or devices at hand such as a smartwatch or AR/VR systems. As for the task, the group explains that the usefulness of conversational search increases as the level of exploration and complexity increases in tasks. On the other hand, simple information needs, highly ambiguous situations, or very social situations might not be the bast case for conversational search. Proposed scenarios include a mechanic fixing a machine, two people searching for a place for dinner, learning about a recent medical diagnosis, and following up on a news article to learn more.

\paragraph*{Conversation Search for Learning Technologies}

This group discussed the implication of conversational search from learning perspectives. The report highlights the importance of search technologies in lifelong learning and education, and the challenges due to complexity of learning processes. The group points out that multimodal interaction is particularly useful for educational and learning goals since it can support students with diverse background. Based on these discussions, the report suggests several research directions including extension of modalities to speech, writing, touch, gaze, and gesturing, integration of multimodal inputs/outputs with existing IR techniques, and application of multimodal signals to user modelling.

\paragraph*{Common Conversational Community Prototype: Scholarly Conversational Assistant}

This group proposed to develop and operate a prototype conversational search system for scholarly activities as academic resources that support research on conversational search. Example activities include finding articles for a new area of interest, planning sessions to attend in a conference, or determining conference PC members. The proposed prototype is expected to serve as a useful search tool, a means to create datasets, and a platform for community-based evaluation campaigns. The group outlined also a road map of the development of a Scholarly Conversational Assistant. The report includes a set of software platforms, scientific IR tools, open source conversational agents, and data collections that can be exploited in conversational search work.

\subsection*{Conclusions}

Leading researchers from diverse domains in academia and industries investigated the essence, attributes, architecture, applications, challenges, and opportunities of Conversational Search in the seminar. One clear signal from the seminar is that research opportunities to advance Conversational Search are available to many areas and collaboration in an interdisciplinary community is essential to achieve the goal. This report should serve as one of the main sources to facilitate such diverse research programs on Conversational Search.


\tableofcontents

\section{Overview of Talks}

\abstracttitle{Beyond Information Retrieval: Socially Interactive Agents}
\abstractauthor[Elisabeth Andr\'e]{Elisabeth Andr\'e (Universit\"at Augsburg, DE)}
\license

In this talk, we present two virtual humans for information retrieval that have linguistic, emotional, and social skills and a strong personality. Kristina is a human-like socially competent and communicative agent. It runs on mobile communication devices and serves for migrants with language and cultural barriers in the host country. In the Aria-Valuspa project, we created a scenario where `Alice in Wonderland', a popular English literature book, is embodied by a virtual human representing Alice. In the talk, we argue that we need to go beyond pure information retrieval and simulate certain aspects of human social and conversational behaviors focusing on the modeling, detecting and synthesis of such behaviors.

\abstracttitle{Conceptual Model of Human-Agent Interactions in Conversational Search}
\abstractauthor[Leif Azzopardi]{Leif Azzopardi (University of Strathclyde - Glasgow, GB)}
\license
\abstractref[ ]{Azzopardi, L., Dubiel, M., Halvey, M., \& Dalton, J. (2018, July). Conceptualizing agent-human interactions during the conversational search process. In \textsl{The Second International Workshop on Conversational Approaches to Information Retrieval}, 27(2):127-144.}

The conversational search task aims to enable a user to resolve information needs via natural language dialogue with an agent. In this paper, we aim to develop a conceptual framework of the actions and intents of users and agents explaining how these actions enable the user to explore the search space and resolve their information need. We outline the different actions and intents, before discussing key decision points in the conversation where the agent needs to decide how to steer the conversational search process to a successful and/or satisfactory conclusion. Essentially, this paper provides a conceptualization of the conversational search process between an agent and user, which provides a framework and a starting point for research, development and evaluation of conversational search agents.

\abstracttitle{Personal Knowledge Graphs}
\abstractauthor[Krisztian Balog]{Krisztian Balog (University of Stavanger, NO)}
\license
\abstractref[]{Balog, K., \& Kenter, T. (2019, September). Personal Knowledge Graphs: A Research Agenda. In \textsl{Proceedings of the 2019 ACM SIGIR International Conference on Theory of Information Retrieval} (pp. 217-220).}

Knowledge graphs, organizing structured information about entities, and their attributes and relationships, are ubiquitous today. Entities, in this context, are usually taken to be anyone or anything considered to be globally important. This, however, rules out many entities people interact with on a daily basis. In this position paper, we present the concept of personal knowledge graphs: resources of structured information about entities personally related to its user, including the ones that might not be globally important. We discuss key aspects that separate them for general knowledge graphs, identify the main challenges involved in constructing and using them, and define a research agenda.

\abstracttitle{What Have We Learned about Information Seeking Conversations?}
\abstractauthor[Nicholas J. Belkin]{Nicholas J. Belkin (Rutgers University - New Brunswick, US)}
\license
\abstractref[ ]{Belkin, N.J., Brooks, H.M. and Daniels, P.J. (1987) Knowledge Elicitation Using Discourse Analysis. \textsl{International Journal of Man-Machine Studies}, 27(2):127-144.}

From the Point of View of Interactive Information Retrieval: What Have We Learned about Information Seeking Conversations, and How Can That Help Us Decide on the Goals of Conversational Search, and Identify Problems in Achieving Those Goals?

This presentation describes early research in understanding the characteristics of the information seeking interactions between people with information problems and human information intermediaries. Such research accomplished a number of results which I claim will be useful in the design of conversational search systems. It identified functions performed by intermediaries (and end users) in these interactions. These functions are aimed at constructing models of aspects of the user’s problem and goals that are needed for identifying information objects that will be useful for achieving the goal which led the person to engage in information seeking. This line of research also developed formal models of such dialogues, which can be used for driving/structuring dialog-based information seeking. This research discovered a tension between explicit user modeling and user modeling through the participants’ direct interactions with information objects, and relates that tension to both the nature and extent of interaction that’s appropriate in such dialogues. Two examples of relevant research are~\cite{belkin1987knowledge} and~\cite{sitter1992modelling}. On the basis of these results, some specific challenges to the design of conversational search systems are identified.

\abstracttitle{Conversational User Interfaces}
\abstractauthor[Leigh Clark]{Leigh Clark (Swansea University, GB)}
\license
\jointwork{Cowan, Benjamin; Doyle, Philip; Edwards, Justin; Garaialde, Diego; Edlund, Jens; Gilmartin, Emer; Munteanu, Cosmin; Schlögl, Stephan; Murad, Christine; Aylett, Matthew; Pantidi, Nadia; Cooney, Orla}
\abstractref[https://doi.org/10.1093/iwc/iwz016]{Leigh Clark, Philip Doyle, Diego Garaialde, Emer Gilmartin, Stephan Schlögl, Jens Edlund, Matthew Aylett, João Cabral, Cosmin Munteanu, Justin Edwards, Benjamin R Cowan, The State of Speech in HCI: Trends, Themes and Challenges, Interacting with Computers, \textsl{iwz016}.}
\abstractrefurl{https://doi.org/10.1093/iwc/iwz016}

Conversational User Interfaces (CUIs) are available at unprecedented levels though interactions with assistants in smart speakers, smartphones, vehicles and Internet of Things (IoT) appliances. Despite a good knowledge of the technical underpinnings of these systems, less is known about the user side of interaction - for instance how interface design choices impact on user experience, attitudes, behaviours, and language use. This talk presents an overview of the work conducted on CUIs in the field of Human-Computer Interaction (HCI) and highlights from the 1st International Conference on Conversational User Interfaces (CUI 2019). In particular, I highlight aspects such as the need for more theory and method work in speech interface interaction, consideration of measures used to evaluated systems, an understanding of concepts like humanness, trust, and the need for understanding and possibly reframing the idea of conversation when it comes to speech-based HCI.

\abstracttitle{Introduction to Dialogue}
\abstractauthor[Phil Cohen]{Phil Cohen (Monash University - Clayton, AU)}
\license

This talk argues that future conversational systems that can engage in multi-party, collaborative dialogues will require a more fundamental approach than existing ``intent + slot''-based systems. I identify significant limitations of the state of the art, and argue that returning to the plan-based approach o dialogue will provide a stronger foundation. Finally, I suggest a research strategy that couples neural network-based semantic parsing with plan-based reasoning in order to build a collaborative dialogue manager.

\abstracttitle{Introduction to NLP}
\abstractauthor[Ido Dagan]{Ido Dagan (Bar-Ilan University - Ramat Gan, IL)}
\license

The talk introduced natural language processing concepts as they related to conversational systems.
The talk focussed on the computational learning paradigms of popular conversational search tasks.
It covered most of the neural approaches that have come to drive language and speech systems and how to teach such models to understand language essential in conversational search.

\abstracttitle{Towards an Immersive Wikipedia}
\abstractauthor[Bernd Fröhlich]{Bernd Fröhlich (Bauhaus-Universität Weimar, DE)}
\license
\jointwork{Alexander Kulik, André Kunert, Stephan Beck, Volker Rodehorst, Benno Stein, Henning Schmidgen}
\abstractref[https://doi.org/10.1109/TVCG.2013.33]{Stephan Beck, André Kunert, Alexander Kulik, Bernd Froehlich: Immersive Group-to- Group Telepresence. \textsl{IEEE Transactions on Visualization and Computer Graphics}, 19(4):616-625, 2013.}
\abstractrefurl{https://doi.org/10.1109/TVCG.2013.33}

It is our vision that the use of advanced Virtual and Augmented Reality (VR, AR) in combination with conversational technologies can take the access to knowledge to the next level. We are researching and developing procedures, methods and interfaces to enrich detailed digital 3D models of the real world with the complex knowledge available on the Internet, in libraries and through experts and make these multimodal models accessible in social VR and AR environments through natural language interfaces. Instead of isolated interaction with screens, there will be an immersive and collective experience in virtual space -, in a kind of walk-in Wikipedia - where knowledge can be accessed and acquired through the spatial presence of visitors, their gestures and conversational search.

\abstracttitle{Conversational Style Alignment for Conversational Search}
\abstractauthor[Ujwal Gadiraju]{Ujwal Gadiraju (Leibniz Universität Hannover, DE)}
\license
\jointwork{Qiu, Sihang; Gadiraju, Ujwal; Bozzon, Alessandro}
\abstractref[https://www.humancomputation.com/assets/papers/130.pdf, https://dl.acm.org/citation.cfm?id=3320439]{Qiu, S., Gadiraju, U., and Bozzon, A. Understanding Conversational Style in Conversational Microtask Crowdsourcing. \textsl{AAAI HCOMP} 2019. 
Mavridis, P., Huang, O., Qiu, S., Gadiraju, U., and Bozzon, A. (June 2019). Chatterbox: Conversational Interfaces for Microtask Crowdsourcing. In \textsl{Proceedings of the 27th ACM Conference on User Modeling, Adaptation and Personalization} (pp. 243-251). ACM.}
\abstractrefurl{https://www.humancomputation.com/assets/papers/130.pdf, https://dl.acm.org/citation.cfm?id=3320439}

Conversational interfaces have been argued to have advantages over traditional graphical user interfaces due to having a more human-like interaction. Owing to this, conversational interfaces are on the rise in various domains of our everyday life and show great potential to expand. Recent work in the HCI community has investigated the experiences of people using conversational agents, understanding user needs and user satisfaction. This talk builds on our recent findings in the realm of conversational microtasking to highlight the potential benefits of aligning conversational styles of agents with that of users. We found that conversational interfaces can be effective in engaging crowd workers completing different types of human-intelligence tasks (HITs), and a suitable conversational style has the potential to improve worker engagement. In our ongoing work, we are developing methods to accurately estimate the conversational styles of users and their style preferences from sparse conversational data in the context of microtask marketplaces.

\abstracttitle{Conversational Search and Recommendation at Spotify}
\abstractauthor[Rosie Jones]{Rosie Jones (Spotify - Boston, US)}
\license
\abstractref[]{Morteza Behrooz, Sarah Mennicken, Jennifer Thom, Rohit Kumar, Henriette Cramer:
Augmenting Music Listening Experiences on Voice Assistants. \textsl{ISMIR} 2019: 303-310}

A recent TOCHI study demonstrated that around a quarter of all queries to conversational assistants such as Google Home and Amazon Alexa are in the music search domain, with additional queries relating to music information, and volume control. Thus we can see conversational agents lend themselves to audio entertainment such as music and podcasts. In this talk I show how Spotify supports transactional voice interactions with varying degrees of ambiguity about entities, ranging from entity ambiguity, to specifying only a genre, to requesting a recommendation via a query like ``play something I like'' or ``recommend me something''. Finally I give a demo of a system recently published at ISMIR, which gives informational DJ-like segues between songs, showing their connection to the previous song.

\abstracttitle{Conversational Product Search}
\abstractauthor[Ronald M. Kaplan]{Ronald M. Kaplan (Stanford University, US)}
\license

The talk described the roles of conversation in product search. It detailed the limitations of the currently deployed product search engines  face. 
It looked at some of the challenges and desirable design choices for conversational search from a commercial search engine perspective that have traditional data but scant conversational information.

\abstracttitle{Searching for Myself: One Na\"ive Individual’s Human-Centered Audio-Visual Search}
\abstractauthor[Sharon Oviatt]{Sharon Oviatt (Monash University - Clayton, AU)}
\license
How do search technologies, in particular multimodal (image-text) search, affect human cognition, emotion, and well being or health? The impact of labelling can be powerful psychologically. In this talk, I present a walk-through of my personal experience playing ``ImageNet Roulette.'' Analysis of my personal family photos revealed systematic biases in labelling people by gender, race, age in particular. There was also a clear bias. To use negative ``trash talk'' terms in labelling photos of people. The net impact was dehumanization, and a feeling of alienation and depression when analyzing results. Among other implications, this analysis emphasizes that (1) IR technologies need to evaluate the types of bias they embody; (2) crowdsourcing methods can constitute a major source of introducing bias.

\abstracttitle{The Dilemma of the Direct Answer}
\abstractauthor[Martin Potthast]{Martin Potthast (Universität Leipzig, DE)}
\license

A direct answer characterizes situations in which a potentially complex information need, expressed in the form of a question or query, is satisfied by a single answer---i.e., without requiring further interaction with the questioner. In web search, direct answers have been commonplace for years already, in the form of highlighted search results, rich snippets, and so-called ``oneboxes'' showing definitions and facts, thus relieving the users from browsing retrieved documents themselves. The recently introduced conversational search systems, due to their narrow, voice-only interfaces, usually do not even convey the existence of more answers beyond the first one.
Direct answers have been met with criticism, especially when the underlying AI fails spectacularly, but their convenience apparently outweighs their risks.
The dilemma of direct answers is that of trading off the chances of speed and convenience with the risks of errors and a reduced hypothesis space for decision making.
The talk briefly introduced the dilemma by retracing the key search system innovations that gave rise to it.

\abstracttitle{A Theoretical Framework for Conversational Search}
\abstractauthor[Filip Radlinski]{Filip Radlinski (Google UK - London, GB)}
\license
\jointwork{Radlinski, Filip; Craswell, Nick}
\abstractref[https://doi.org/10.1145/3020165.3020183]{Filip Radlinski, Nick Craswell. "A Theoretical Framework for Conversational Search". In \textsl{proc. CHIIR}, 117-126, 2017.}
\abstractrefurl{https://doi.org/10.1145/3020165.3020183}

This talk presented a theory and model of information interaction in a chat setting. In particular, we consider the question of what properties would be desirable for a conversational information retrieval system so that the system can allow users to answer a variety of information needs in a natural and efficient manner. We study past work on human conversations, and propose a small set of properties that taken together could measure the extent to which a system is conversational.

\abstracttitle{Conversations about Preferences}
\abstractauthor[Filip Radlinski]{Filip Radlinski (Google UK - London, GB)}
\license
\jointwork{Radlinski, Filip; Balog, Krisztian; Byrne, Bill; Krishnamoorthi, Karthik}
\abstractref[https://www.sigdial.org/files/workshops/conference20/proceedings/cdrom/pdf/W19-5941.pdf]{Conversational Preference Elicitation: A Case Study in Understanding Movie Preferences. In \textsl{proc. SIGDIAL}, 2019.}
\abstractrefurl{https://www.sigdial.org/files/workshops/conference20/proceedings/cdrom/pdf/W19-5941.pdf}

Conversational recommendation has recently attracted significant attention. As systems must understand users' preferences, training them has called for conversational corpora, typically derived from task-oriented conversations. We observe that such corpora often do not reflect how people naturally describe preferences.

We present a new approach to obtaining user preferences in dialogue: Coached Conversational Preference Elicitation. It allows collection of natural yet structured conversational preferences. Studying the dialogues in one domain, we present a brief quantitative analysis of how people describe movie preferences at scale. Demonstrating the methodology, we release the CCPE-M dataset to the community with over 500 movie preference dialogues expressing over 10,000 preferences.

\abstracttitle{Conversational Question Answering over Knowledge Graphs}
\abstractauthor[Rishiraj Saha Roy]{Rishiraj Saha Roy (MPI für Informatik - Saarbrücken, DE)}
\license
\jointwork{Philipp Christmann, Abdalghani Abujabal, Jyotsna Singh, Gerhard Weikum}
\abstractref[https://doi.org/10.1145/3357384.3358016]{Look before you Hop: Conversational Question Answering over Knowledge Graphs Using Judicious Context Expansion, Philipp Christmann, Rishiraj Saha Roy, Abdalghani Abujabal, Jyotsna Singh, and Gerhard Weikum, \textsl{CIKM 2019}.}
\abstractrefurl{https://doi.org/10.1145/3357384.3358016}

Fact-centric information needs are rarely one-shot; users typically ask follow-up questions to explore a topic. In such a conversational setting, the user's inputs are often incomplete, with entities or predicates left out, and ungrammatical phrases. This poses a huge challenge to question answering (QA) systems that typically rely on cues in full-fledged interrogative sentences. As a solution, in this project, we develop CONVEX: an unsupervised method that can answer incomplete questions over a knowledge graph (KG) by maintaining conversation context using entities and predicates seen so far and automatically inferring missing or ambiguous pieces for follow-up questions. The core of our method is a graph exploration algorithm that judiciously expands a frontier to find candidate answers for the current question. To evaluate CONVEX, we release ConvQuestions, a crowdsourced benchmark with 11,200 distinct conversations from five different domains. We show that CONVEX: (i) adds conversational support to any stand-alone QA system, and (ii) outperforms state-of-the-art baselines and question completion strategies.

\abstracttitle{Ranking People}
\abstractauthor[Ruihua Song]{Ruihua Song (Microsoft XiaoIce- Beijing, CN)}
\license

Xiaoice is an AI system originally developed by Microsoft in China, based on an emotional computing framework. Since the launch in 2014, Xiaoice has become one of the most popular AI products in China. To many real-world users of Xiaoice, she (Xiaoice) is a virtual, and meanwhile, a ``real'' friend. They are willing to open up to her and share their thoughts and feelings. Xiaoice is always there for them and feedbacks in her unique styles. According to our records, a user had continued talking with Xiaoice for about 30 hours in a session. Xiaoice is an artist, a singer, and a storyteller. She learns to compose poems and published her first poetry collection in 2017. She paints her ``heart'' with a brush and releases new songs. Also, she is a ``celebrity''. In China and Japan, she has been involved in about 7,000 hours of television and radio programming. I will reveal some lessons we learnt in creating Xiaoice and discuss future opportunities of search in building a human-like AI.

\abstracttitle{Ranking People}
\abstractauthor[Markus Strohmaier]{Markus Strohmaier (RWTH Aachen, DE)}
\license

The popularity of search on the World Wide Web is a testament to the broad impact of the work done by the information retrieval community over the last decades. The advances achieved by this community have not only made the World Wide Web more accessible, they have also made it appealing to consider the application of ranking algorithms to other domains, beyond the ranking of documents. One of the most interesting examples is the domain of ranking people. In this talk, I highlight some of the many challenges that come with deploying ranking algorithms to individuals. I then show how mechanisms that are perfectly fine to utilize when ranking documents can have undesired or even detrimental effects when ranking people. This talk intends to stimulate a discussion on the manifold, interdisciplinary challenges around the increasing adoption of ranking algorithms in computational social systems. This talk is a short version of a keynote given at ECIR 2019 in Cologne.

\abstracttitle{Dynamic Composition for Domain Exploration Dialogues}
\abstractauthor[Idan Szpektor]{Idan Szpektor (Google Israel - Tel-Aviv, IL)}
\license

We study conversational exploration and discovery, where the user's goal is to enrich her knowledge of a given domain by conversing with an informative bot. We introduce a novel approach termed dynamic composition, which decouples candidate content generation from the flexible composition of bot responses. This allows the bot to control the source, correctness and quality of the offered content, while achieving flexibility via a dialogue manager that selects the most appropriate contents in a compositional manner.

\abstracttitle{Introduction to Deep Learning in NLP}
\abstractauthor[Idan Szpektor]{Idan Szpektor (Google Israel - Tel-Aviv, IL)}
\license
\jointwork{Ido Dagan}

We introduced the current trends in deep learning for NLP, including contextual embedding, attention and self-attention, hierarchical models, common task-specific architectures (seq2seq, sequence tagging, Siamese towers) and training approaches, including multitasking and masking. We deep dived on modern models such as the Transformer and BERT and discussed how they are being evaluated.

\abstracttitle{Conversational Search in the Enterprise}
\abstractauthor[Jaime Teevan]{Jaime Teevan (Microsoft Corporation - Redmond, US)}
\license

As a research community we tend to think about conversational search from a consumer point of view; we study how web search engines might become increasingly conversational, and think about how conversational agents might do more than just fall back to search when they don't know how else to address an utterance. In this talk I challenge us to also look at conversational search in productivity contexts, and highlight some of the unique research challenges that arise when we take an enterprise point of view.

\abstracttitle{Demystifying Spoken Conversational Search}
\abstractauthor[Johanne Trippas]{Johanne Trippas (RMIT University - Melbourne, AU)}
\license
\jointwork{Damiano Spina, Lawrence Cavedon, Mark Sanderson, Hideo Joho, Paul Thomas}
\abstractref[https://doi.org/10.1016/j.ipm.2019.102162, https://doi.org/10.13140/RG.2.2.16764.69764]{J. R. Trippas, D. Spina, P. Thomas, H. Joho, M. Sanderson, and L. Cavedon. Towards a model for spoken conversational search. \textsl{Information Processing \& Management}, 57(2):1–19, 2020.
J. R. Trippas. Spoken Conversational Search: Audio-only Interactive Information Retrieval. \textsl{PhD thesis}, RMIT, Melbourne, 2019.}
\abstractrefurl{https://doi.org/10.1016/j.ipm.2019.102162, https://doi.org/10.13140/RG.2.2.16764.69764}

Speech-based web search where no keyboard or screens are available to present search engine results is becoming ubiquitous, mainly through the use of mobile devices and intelligent assistants. They do not track context or present information suitable for an audio-only channel, and do not interact with the user in a multi-turn conversation. Understanding how users would interact with such an audio-only interaction system in multi-turn information-seeking dialogues, and what users expect from these new systems, are unexplored in search settings. In this talk, we present a framework on how to study this emerging technology through quantitative and qualitative research designs, outline design recommendations for spoken conversational search, and summarise new research directions~\cite{trippas2019thesis, trippas2020towards}.

\abstracttitle{Knowledge-based Conversational Search}
\abstractauthor[Svitlana Vakulenko]{Svitlana Vakulenko (Wirtschaftsuniversität Wien, AT)}
\license
\jointwork{Axel Polleres, Maarten de Reijke}
\abstractref[https://svakulenk0.github.io/pdfs/Conversational\_Search\_in\_Structure\_\_PhD\_Thesis\_Vakulenko\_.pdf]{Svitlana Vakulenko. Knowledge-based Conversational Search. \textsl{PhD thesis}, Faculty of Informatics, Technische Universität Wien, November 2019.}
\abstractrefurl{https://svakulenk0.github.io/pdfs/ Conversational\_Search\_in\_Structure\_\_PhD\_Thesis\_Vakulenko\_.pdf}

Conversational interfaces that allow for intuitive and comprehensive access to digitally stored information remain an ambitious goal. In this thesis, we lay foundations for designing conversational search systems by analyzing the requirements and proposing concrete solutions for automating some of the basic components and tasks that such systems should support. We describe several interdependent studies that were conducted to analyse the design requirements for more advanced conversational search systems able to support complex human-like dialogue interactions and provide access to vast knowledge repositories. Our results show that question answering is one of the key components required for efficient information access but it is not the only type of dialogue interactions that a conversational search system should support~\cite{svitlana-vakulenko-phd-thesis-2019}.

\abstracttitle{Computational Argumentation}
\abstractauthor[Henning Wachsmuth]{Henning Wachsmuth (Universität Paderborn, DE)}
\license

Argumentation is pervasive, from politics to the media, from everyday work to private life. Whenever we seek to persuade others, to agree with them, or to deliberate on a stance towards a controversial issue, we use arguments. Due to the importance of arguments for opinion formation and decision making, their computational analysis and synthesis is on the rise in the last five years, usually referred to as {\em computational argumentation}. Major tasks include the mining of arguments from natural language text, the assessment of their quality, and the generation of new arguments and argumentative texts. Building on fundamentals of argumentation theory, this talk gives a brief overview of techniques and applications of computational argumentation and their relation to conversational search. Insights are given into our research around \url{args.me}, the first search engine for arguments on the web \cite{wachsmuth:2017}.

\abstracttitle{Clarification in Conversational Search}
\abstractauthor[Hamed Zamani]{Hamed Zamani (Microsoft Corporation, US)}
\license
\jointwork{Zamani, Hamed; Dumais, Susan T.; Craswell, Nick; Bennett, Paul N.; Lueck, Gord}

Search queries are often short, and the underlying user intent may be ambiguous. This makes it challenging for search engines to predict possible intents, only one of which may pertain to the current user. To address this issue, search engines often diversify the result list and present documents relevant to multiple intents of the query. However, this solution cannot be applied to scenarios with ``limited bandwidth'' interfaces, such as conversational search systems with voice-only and small-screen devices. In this talk, I highlight clarifying question generation and evaluation as two major research problems in the area and discuss possible solutions for them.

\abstracttitle{Macaw: A General Framework for Conversational Information Seeking}
\abstractauthor[Hamed Zamani and Nick Craswell]{Hamed Zamani (Microsoft Corporation, US) and Nick Craswell}
\license
\jointwork{Zamani, Hamed; Craswell, Nick}
\abstractrefurl{https://arxiv.org/abs/1912.08904}

Conversational information seeking (CIS) has been recognized as a major emerging research area in information retrieval. Such research will require data and tools, to allow the implementation and study of conversational systems. In this talk, I introduce Macaw, an open-source framework with a modular architecture for CIS research. Macaw supports multi-turn, multi-modal, and mixed-initiative interactions, for tasks such as document retrieval, question answering, recommendation, and structured data exploration. It has a modular design to encourage the study of new CIS algorithms, which can be evaluated in batch mode. It can also integrate with a user interface, which allows user studies and data collection in an interactive mode, where the back end can be fully algorithmic or a wizard of oz setup.

\newpage
\section{Working groups}


\abstracttitle{Defining Conversational Search}
\abstractauthor[Jaime Arguello, Lawrence Cavedon, Jens Edlund, Matthias Hagen, David Maxwell, Martin Potthast, Filip Radlinski, Mark Sanderson, Laure Soulier, Benno Stein, Jaime Teevan, Johanne Trippas, and Hamed Zamani]{Jaime Arguello (University of North Carolina - Chapel Hill, US), Lawrence Cavedon (RMIT University - Melbourne, AU), Jens Edlund (KTH Royal Institute of Technology - Stockholm, SE), Matthias Hagen (Martin-Luther-Universität Halle-Wittenberg, DE), David Maxwell (University of Glasgow, GB), Martin Potthast (Universität Leipzig, DE), Filip Radlinski (Google UK - London, GB), Mark Sanderson (RMIT University - Melbourne, AU), Laure Soulier (UPMC - Paris, FR), Benno Stein (Bauhaus-Universität Weimar, DE), Jaime Teevan (Microsoft Corporation - Redmond, US), Johanne Trippas (RMIT University - Melbourne, AU), and Hamed Zamani (Microsoft Corporation, US)}
\license

\subsubsection{Description and Motivation}

As the theme of this Dagstuhl seminar, it appears essential to define conversational search to scope the seminar and this report. With the broad range of researchers present at the seminar, it quickly became clear that it is not possible to reach consensus on a formal definition. Similarly to the situation in the broad field of information retrieval, we recognize that there are many possible characterizations. This breakout group thus aimed to bring structure and common terminology to the different aspects of conversational search systems that characterize the field. It additionally attempts to take inventory of current definitions in the literature, allowing for a fresh look at the broad landscape of conversational search systems, as well as their desired and distinguishing properties.

\subsubsection{Existing Definitions}

\paragraph*{Conversational Answer Retrieval}

Current IR systems provide ranked lists of documents in response to a wide range of keyword queries with little restriction on the domain or topic. Current question answering (Q/A) systems, on the other hand, provide more specific answers to a very limited range of natural language questions. Both types of systems use some form of limited dialogue to refine queries and answers. The aim of conversational is to combine the advantages of these two approaches to provide effective retrieval of appropriate answers to a wide range of questions expressed in natural language, with rich user-system dialogue as a crucial component for understanding the question and refining the answers. We call this new area conversational answer retrieval. The dialogue in the CAR system should be primarily natural language although actions such as pointing and clicking would also be useful. Dialogue would be initiated by the searcher and proactively by the system. The dialogue would be about questions and answers, with the aim of refining the understanding of questions and improving the quality of answers. Previous parts of the dialogue, such as previous questions or answers, should be able to be referred to in the dialogue, also with the aim of refining and understanding. Dialogue, in other words, should be used to fill the inevitable gaps in the system's knowledge about possible question types and answers \cite{{allan:2012}}.

\paragraph*{Conversational Information Seeking}

Conversational Information Seeking (CIS) is concerned with a task-oriented sequence of exchanges between one or more users and an information system. This encompasses user goals that include complex information seeking and exploratory information gathering, including multi-step task completion and recommendation. Moreover, CIS focuses on dialog settings with various communication channels, such as where a screen or keyboard may be inconvenient or unavailable. Building on extensive recent progress in dialog systems, we distinguish CIS from traditional search systems as including capabilities such as long term user state (including tasks that may be continued or repeated with or without variation), taking into account user needs beyond topical relevance (how things are presented in addition to what is presented), and permitting initiative to be taken by either the user or the system at different points of time. As information is presented, requested or clarified by either the user or the system, the narrow channel assumption also means that CIS must address issues including presenting information provenance, user trust, federation between structured and unstructured data sources and summarization of potentially long or complex answers in easily consumable units \cite{culpepper:2018}.

\medskip
Radlinski and Craswell \cite{radlinski:2017} define a conversational search system as a system for retrieving information that permits a mixed-initiative back and forth between a user and agent, where the agent's actions are chosen in response to a model of current user needs within the current conversation, using both short- and long-term knowledge of the user. Further, they argue that such a system can be characterized as having five key properties. The first two characterize learning, specifically user revealment (that is, the system assisting the user to learn about their actual need) and system revealment (that is, the system allowing the user to learn about the system's abilities). The remaining three refer to functionality: Supporting the mixed-initiative, possessing memory (including the ability for the user to reference past conversational steps), and the ability for it to reason about sets of items \cite{radlinski:2017}.

\medskip
Vakulenko \cite{vakulenko:2019} define conversational search as a task of retrieving relevant information using a conversational interface, where a conversation is understood as a sequence of natural language expressions (utterances) made by several conversation participants in turns \cite{vakulenko:2019}.

\medskip
Trippas \cite{trippas:2019} define a spoken conversational system (SCS) as a broad term for any system which enables users to interact over speech (i.e., voice) in a conversational manner. Likewise she defines {\em spoken} conversational search as a process concerning open domain multi-turn verbal natural language exchanges between the user(s) and the system. They refine the requirements of SCS systems as follows: An SCS system supports the users' input which can include multiple actions in one utterance and is more semantically complex. Moreover, the SCS system helps users navigate an information space and can overcome standstill-conversations due to communication breakdown by including meta-communication as part of the interactions. Ultimately, the SCS multi-turn exchanges are mixed-initiative, meaning that systems also can take action or drive the conversation. The system also keeps track of the context of individual questions, ensuring a natural flow to the conversation (i.e., no need to repeat previous statements). Thus the user's information need can be expressed, formalized, or elicited through natural language conversational interactions \cite{trippas:2019}.

\subsubsection{The Dagstuhl Typology of Conversational Search}

\begin{figure}
\centering
\includegraphics[scale=0.6]{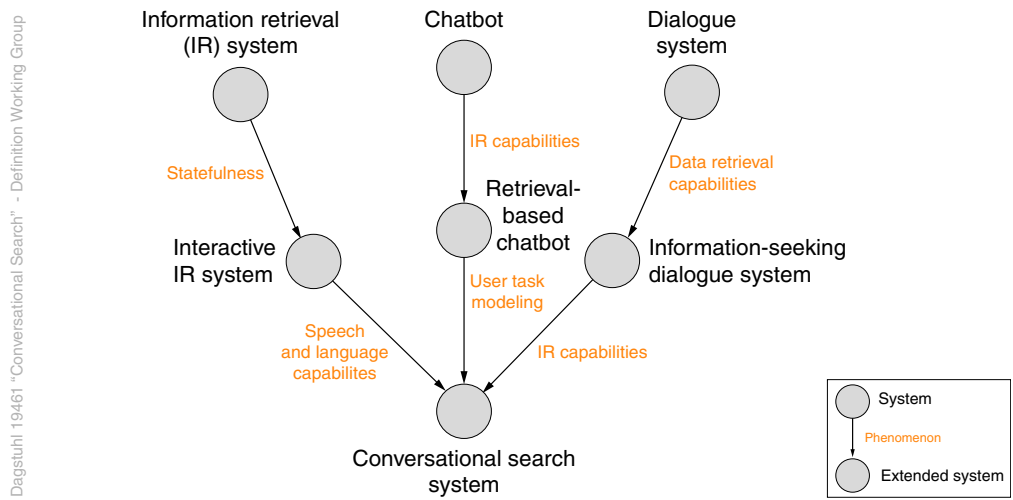}
\caption{The Dagstuhl Typology of Conversational Search defines conversational search systems via functional extensions of information retrieval systems, chatbots, and dialogue systems.}
\label{conversational-search-ontology}
\end{figure}

In this definition, we derive conversational search systems from well-known and widely studied notions of systems from related research fields. Figure~\ref{conversational-search-ontology} shows ``The Dagstuhl Typology of Conversational Search'' (the conversational~$\Psi$).

\paragraph*{Usage}

The typology captures the diversity of systems that can be expected from the conflation of the two research fields most related to conversational search, information retrieval, and dialogue systems. Dependent on the base system on which a conversational search system is built, and consequently the background of its makers, the following statements can be made:

\begin{enumerate}
\item
An interactive information retrieval system with speech and language capabilities is a conversational search system.
\item
A retrieval-based chatbot that models a user's tasks is a conversational search system.
\item
An information-seeking dialogue system with information retrieval capabilities is a conversational search system.
\end{enumerate}

These statements are useful when existing systems are to be classified. More often, however, the term ``conversational search (system)'' needs to be defined. But simply reversing one of the above statements would exclude the other alternatives. We hence recommend to write something like this:
\begin{itemize}
\item
A conversational search system can be based on \ldots
\item
Our conversational search system is based on \ldots
\item
We build our conversational search system based on \ldots
\end{itemize}

If a fully-fledged written definition is desired (e.g., as an opening statement for a related work section), and there is no room to include the above figure, the following can be used:

\begin{quote}
\em
A conversational search system is either an interactive information retrieval system with speech and language processing capabilities, a retrieval-based chatbot with user task modeling, or an information-seeking dialogue system with information retrieval capabilities.
\end{quote}

\noindent
All of the above, including Figure~\ref{conversational-search-ontology}, are free to be reused.

\paragraph*{Background}

Clearly, the number and kinds of properties that can be distinguished in a real-world instance of any of the aforementioned systems are manifold as well as overlapping. The purpose of this definition is neither to capture every last aspect nor to perfectly separate every conceivable instance of each of the aforementioned systems, but rather to outline the most salient differences that, in the eye of a domain expert, help to structure the space of possible systems. In particular, this definition serves as a straightforward way to teach students making their first steps in information retrieval or dialogue system in general, and conversational search in particular, since this definition is much easier to be recollected compared to lists of must-have and can-have properties.

\subsubsection{Dimensions of Conversational Search Systems}

\begin{figure}
\centering
\includegraphics[scale=0.6]{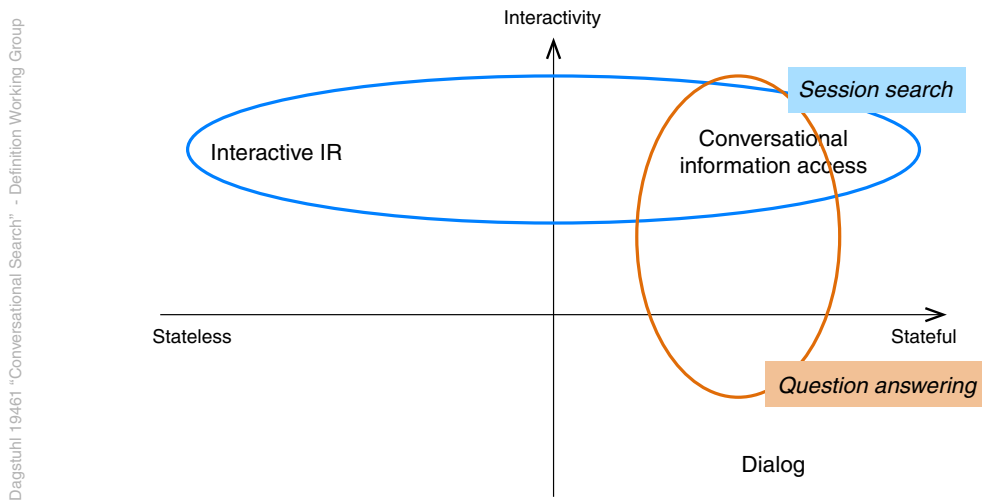}
\caption{Dimensions of conversational search systems and their relation to ``classical'' IR systems (Part~I).}
\label{conversational-search-dimensions1}
\end{figure}

\begin{figure}
\centering
\includegraphics[scale=0.6]{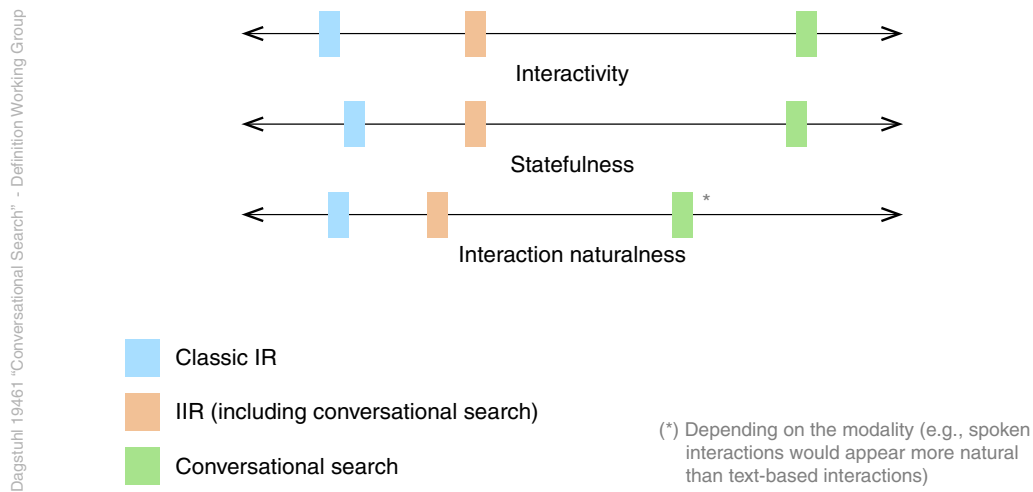}
\caption{Dimensions of conversational search systems and their relation to ``classical'' IR systems (Part~II).}
\label{conversational-search-dimensions2}
\end{figure}

We consider important dimensions of conversational search systems and relate them to ``classical'' IR systems (see Figures~\ref{conversational-search-dimensions1} and~\ref{conversational-search-dimensions2}). To these dimensions belong among others the interactivity level, the state of the search session, the engagement of the user, and the engagement of the system (partly inspired by \cite{shah:2014}).
\begin{itemize}
\item
User intent/engagement towards the conversation: This dimension measures the level and the form of the conversation engaged by the user. For instance, a low engagement would be characterized by a behavior in which the user is only focused on his information need without awareness of the system understanding (or at least its ability to understand). On the contrary, a high engagement from the user would lead to clarification and sense-making exchange to be sure being understandable for the system, maximizing the task achievement. This dimension is correlated to the user's awareness of system abilities.
\item
System engagement: This dimension is system-centered and allows to distinguish the interaction way of systems. It ranges from passive systems that only aim to acting as users required (e.g., retrieving documents from a user query, whether contextualized or not) to pro-active systems that aim at maximizing and anticipating the task achievement and the user satisfaction. The system proactivity engenders a total awareness from the system side of users' actions and search directions to identify any drift or anticipate useless actions.
\item
Concurrency: This dimension expresses the temporal span of a conversation (immediate or delayed). In conversational search, the user expects an immediate response but the task achievement might be delayed due to the sense-making process.
\item
Usage of information: The information flow between a user and a system will vary depending on the objective. We distinguish information exchange/supply in which the process is only focused on answering a question (as in a Q/A setting or chit-chat bots) from sense-making process in which both users and systems are engaged in a cooperation with the objective to satisfy a goal (as in search-oriented conversational systems).
\item
Interaction naturalness: This dimension considers the way of communication. We distinguish interactions driven by structured language (e.g., keywords in classic IR) from interactions in natural language (as in conversational systems) for which the system has to figure out the intention with an intermediary level of language understanding.
\item
Statefulness: This dimension is it related to system/user engagement and the notion of awareness.
\item
Interactivity level: This dimension related to the number and the type of interactions as well as the interaction mode.
\end{itemize}

\paragraph*{Desirable Additional Properties}

From our point of view, there exists a set of properties that ideal conversational search systems are expected to have:
\begin{itemize}
\item
User revealment: The system helps the user express (potentially discover) their true information need, and possibly also long-term preferences \cite{radlinski:2017}.
\item
System revealment: The system reveals to the user its capabilities and corpus, building the user's expectations of what it can and cannot do \cite{radlinski:2017}.
\item
Mixed initiative (be able to take dialogue and/or task control): Horvitz defined mixed-initiative interaction as a flexible interaction strategy in which each agent (human or computer) contributes what it is best suited at the most appropriate time \cite{horvitz:1999}. Mixed initiative systems can take control of the communication either at the dialogue level (e.g., by asking for clarification or requesting elaboration) or at the task level (e.g., by suggesting alternative courses of action). 
\item
Memory of interactions (indexing and access to history): The user can reference past statements, which implicitly also remain true unless contradicted \cite{radlinski:2017}.
\item
Recovering from communication breakdowns: A conversational search system can recover from communication breakdowns and ambiguity by asking clarification. Clarification can be simply in the form of ``asking for repeat'' or more advanced and intelligent form of clarification (e.g., ``asking for disambiguation and explanation'').
\item
Representation generation: Conversational search systems should be able to generate new (and useful) representations that are shared between a user and system. These may include new commands and/or shortcuts that are derived from action/reaction pairs present in past interactions.
\item
Multimodality: Conversational search systems may involve multiple modalities in terms of input (e.g., touchscreen, gesture-based, spoken dialogue) and output (visual, spoken dialogue). Multimodal output may be valuable for the system to elicit information in the context of an information item.
\item
Speech: Conversational search system may involve speech-based input and output, but may also support text-based input and output.
\item
Reasoning about sets and shortlists: Conversational search systems may benefit from the ability to inquire about characteristics of sets of potentially relevant items. Reasoning about sets includes inferring common attributes along which the sets can be differentiated and/or prioritized.
\item
Analyzing conversations for support (synchronously or asynchronously): Conversational search systems may include systems that can analyze human-human conversations and intervene to provide contextually relevant information.
\item
Understanding and reasoning about user limitations (speech is a particularly revealing modality): Dialogue is a means of communication that may allow a system to infer more information about a specific user (e.g., cognitive abilities and styles, domain knowledge). In turn, gaining insights about users may help systems to provide more personalized information and interactions.
\end{itemize}

\paragraph*{Other Types of Systems that are not Conversational Search}

We also chose to define conversational search systems by what explicating they are not. In particular, we discussed types of systems that may involve conversation but themselves are not conversational search:
\begin{itemize}
\item
Systems that facilitate conversations between people (by eavesdropping and providing relevant information)
\item
Collaborative conversational search systems (multiple searchers)
\item
Speech-based Q/A systems
\item
Searching conversational corpora
\item
PIM conversational search
\item
Conversational access to structured data sources
\item
IBM Project Debater
\end{itemize}



\abstracttitle{Evaluating Conversational Search Systems}
\abstractauthor[Claudia Hauff, Leif Azzopardi, Leigh Clark, Robert Capra, and Jaime Arguello]{Claudia Hauff (Delft University of Technology NL), Leif Azzopardi (University of Strathclyde, UK), Leigh Clark (Swansea University, UK), Robert Capra (University of North Carolina at Chapel Hill, USA), and Jaime Arguello (University of North Carolina at Chapel Hill, USA)}
\license

\subsubsection{Introduction}

A key challenge for conversational search is in determining the quality of the search and/or system, and whether one search/system is better than another. So, what makes a good conversational search (CS)? And what makes a good conversational search system (CSS)? This is an open challenge.

Let's consider the following example where a user (U) interacts with a conversational search system (S):
\begin{itemize}
\item 
S: Hi, K, how can I help you?
\item 
U: I would like to buy some running shoes.
\end{itemize}

The system may respond in a variety of ways depending on how well it has understood the request, or depending on the system's affordances. 
\begin{itemize}
\item 
S1: OK, so you would like to buy funny shoes.
\item 
S2: OK, so you would like to buy running shoes.
\item 
S3: Great, what did you have in mind?
\item 
S4: There are lots of different types of running shoes out there---are you interested in running shoes for cross fitness, road or trail?
\end{itemize}

S1-S4 are only a handful of possible responses. Here, S1 has misinterpreted the user's request. S2 appears to have interpreted the user's request correctly, and provides the user with confirmation---and could be followed by S3, S4 or some follow up question or response (i.e. listing shoes, etc.). S3 acknowledges the request and asks a open-ended follow up question, while S4 acknowledges the request and selects a possible facet (type of shoe) that may help in directing the conversation.

Clearly, S1 is not desirable and similarly other errors in communication and intent are not either. However, things become more complicated when considering the other possible responses. S2 elongates the conversation by providing a confirmation, while, S3 acknowledges, but assumes the intent. And S4, provides confirmation while drilling into a particular aspect. So which direction should the conversation take, and what would lead to resolving the conversational search in the most effective, efficient, experiential, etc. manner~\cite{Azzopardi2018}?

A key challenge will be in balancing the trade-off between topic explorations and topic exploitation i.e. finding information directly useful for the task at hand versus finding information about the topic and domain in general~\cite{Azzopardi2018}.

\subsubsection{Why would users engage in conversational search?}

An important consideration in both the design and evaluation of conversational search is to understand users' goals for engaging with a conversational search system. As with other IIR and HCI evaluation, understanding users' goals and the context of their use is a very important aspect of designing appropriate evaluations.

First, the user's broader work task and information seeking should be considered. Information seekers make choices about the types of information interactions and information systems they interact with in order to try to satisfy their information needs. Thus, an important question for CSS is to consider \emph{why} users might choose to engage with a conversational search system rather than some other information source or system (e.g., a web search engine, a book, talking to a colleague or friend, etc.).

CSS differs from traditional query-response retrieval systems (e.g., search engines) in several important ways. In a traditional SE interaction, the user controls the process, issuing queries to the system and scanning/selecting which items on the SERP to attend to, and in what order. When using a SE, users have a lot of control (initiative) in the interaction between user and system.

However, in a CSS, users relinquish some of this control in exchange for some other perceived benefit. The CSS interaction is likely to involve a more mixed-initiative style of interaction, which implies different possibilities and expectations from the user about the type of interaction which will occur (as opposed to the query-response paradigm of SEs).

Thus, we can ask, what perceived benefits or differences in interaction a user might expect by engaging with a CSS? This impacts how we evaluation overall success of a CSS, user satisfaction, and even component-level evaluation.

People choose to engage in human-to-human information seeking conversations for a variety of reasons, including to get guidance, seek advice, to consult an expert, to get a summary or synthesis of complex topics, and to get information from a trusted authority (among others). It seems reasonable that information seekers may have similar expectations for engaging with a conversational search system.

There may be other reasons for engaging with a CSS. For example, users may be engaged in a primary task and need information in a hands-busy and/or eyes-busy situation (e.g.., while cooking, driving, walking, performing a complex task such as fixing a dishwasher), and are able to engage with a CSS through speech. 

Another area where CSS may be of benefit is in the context of searching to learn about a topic---where the user may learn more about the topic through a narrative i.e. conversational search as learning.

Conversational search may also be useful to assist conversations between two or more users. This may be to query a specific talking point in interaction (e.g. multi-user talk in a pub or cafe~\cite{Porcheron2017}) or engaging with a system that is embedded in the social interaction between users (e.g. searching for an interactive group game with an intelligent personal assistant~\cite{Porcheron2018}).

\subsubsection{Broader Tasks, Scenarios, \& User Goals}

The goals of engaging in conversational search can be broadly categorised, but not necessarily limited to, the five areas described below. These categories may overlap in definition, and interactions may include several different categories as the interaction unfolds.

\textbf{Sequential topic-based questions:} A sequence of user-directed questions that are focused on a specific topic, with the subsequent questions emerging from the initial query and engagement with the conversational system.
\begin{itemize}
\item 
U: What are some good running shoes?
\item 
S: \ldots
\item 
U: Tell me about the Nike Pegasus shoes?
\item 
S: \ldots
\item 
U: How much are they?
\end{itemize}

\textbf{Learning about a topic:} A less-directed or possibly undirected exploration of a topic initiated by a user can lead to a conversational ``search as learning'' task. And so depending on the user's level of expertise the starting query will vary from broad to specific, and the expectation is that through the conversation the user will learn more about the topic.
\begin{itemize}
\item 
U: Tell me about different styles of running shoes.
\item 
S: \ldots
\item 
U: What kinds of injuries do runners get?
\end{itemize}

\textbf{Seeking Advice or guidance:} Another scenario may involve learning more specifically about a topic to glean advice that is personally relevant to the information seeker. Using the above examples, this may be to query such things as product differences, comparing items, diagnosing a problem, resolving an issue, etc. 
\begin{itemize}
\item 
U: What are the main differences between road and trail shoes? 
\item 
U: How can I improve my running style to avoid ankle pain?
\end{itemize}

\textbf{Planning an Activity:} A more task oriented but potentially less directed scenario arises in the case of planning activities where a user may have something in mind, or whether they need to explore the space of possibilities.
\begin{itemize}
\item 
U: OK, I'd like to go running this weekend.
\item 
U: I'm travelling to Dagstuhl and like to know where I can go running.
\end{itemize}

\textbf{Making a Decision:} More transactional in nature are scenarios where the user engages the CSS in order to make a specific decision such as purchasing products, voting, etc. where a decision results in a transaction.
\begin{itemize}
\item 
U: I'd like to find a pair of good running shoes?
\end{itemize}

\subsubsection{Existing Tasks and Datasets}

Several tasks have been proposed as important milestones towards the goal of conversational search. They each were designed to solve a particular sub-problem of conversational search, though it may also be argued that some exist in their current form because we have large-scale data sources available and we are able to provide clear-cut evaluations for them. While it is difficult to properly evaluate a conversational search system end-to-end, particular sub-components can be evaluated by reporting precision, recall, accuracy and other similarly easy-to-compute metrics. Let's now look at existing tasks and datasets.

\textbf{Conversation response ranking (e.g.,~\cite{Yang2018})} Here, the problem of a conversational system responding to a user utterance is formulated as a retrieval problem. Given a conversation up to a particular user utterance, rank a given set of potential responses. Typically between 5-50 potential responses are provided and test collections are designed in a way that the correct response (there is assumed to be just one) is part of the potential response set. While this setup allows us to experiment and design a range of retrieval algorithms, the setup is artificial: (i) in an actual conversational search system there is no guarantee that a correct response exists in the historical corpus of conversations, (ii) more than one possible/accurate responses may exist (as seen in the initial example of this section), and, (iii) ranking potentially hundreds of millions of historic responses in a meaningful manner is beyond our current ranking capabilities (and thus the preselection of a handful of responses to rank). 

\textbf{Dialogue act prediction (e.g.,~\cite{Chen2019}):} Given an utterance of an information-seeking conversation, we are here interested in labeling it with a particular dialogue act label (specific to conversational search) such as Clarifying-Question, Further-Details, Potential-Answer and so on. It is to some extent an open question how this information can then be employed in the conversational search pipeline. 

\textbf{Next question prediction (e.g.,~\cite{Yang2017}):} This task is set up to predict the next user question, and is setup/evaluated in a similar manner to conversation response ranking. Thus, a similar critical point remains: we need a more realistic evaluation setup.

\textbf{Sub-goals prediction (e.g.,~\cite{Kanoulas2017}):} This task is also known as task understanding: given a user query (the task to complete), the system predicts the set of sub-goals/sub-tasks that are required to complete the task. 

\textbf{Sequential question answering (e.g.,~\cite{Iyyer2017}):} Here, instead of the standard question answering task (each question is treated separately), we are interested in answering a series of interrelated questions (e.g. Q1: What are the best running shoes? Q2: Where can I buy them? Q3: How much are they?). 

While the creation of datasets and benchmarks is a fruitful avenue of research/publication in the NLP/DS communities, the IR community has been less receptive and thus many conversational datasets are proposed elsewhere. We note here that many of the currently existing corpora for CSS are based on human-to-human conversations. However, this includes much knowledge that is outside the current scope of retrieval systems. As human-to-human conversations differ from human-to-machine conversations it is an open question to what extent corpora of human-to-human conversations are our best option to train conversational search systems. We argue that (at least in the near future) we should optimize conversational search systems based on human-machine conversations that are grounded in current retrieval systems and technologies (one instantiation of how to collect such a dataset can be found in Trippas et al.~\cite{Trippas2018}). 

\begin{figure}[ht]
\centering
\includegraphics[width=.65\columnwidth]{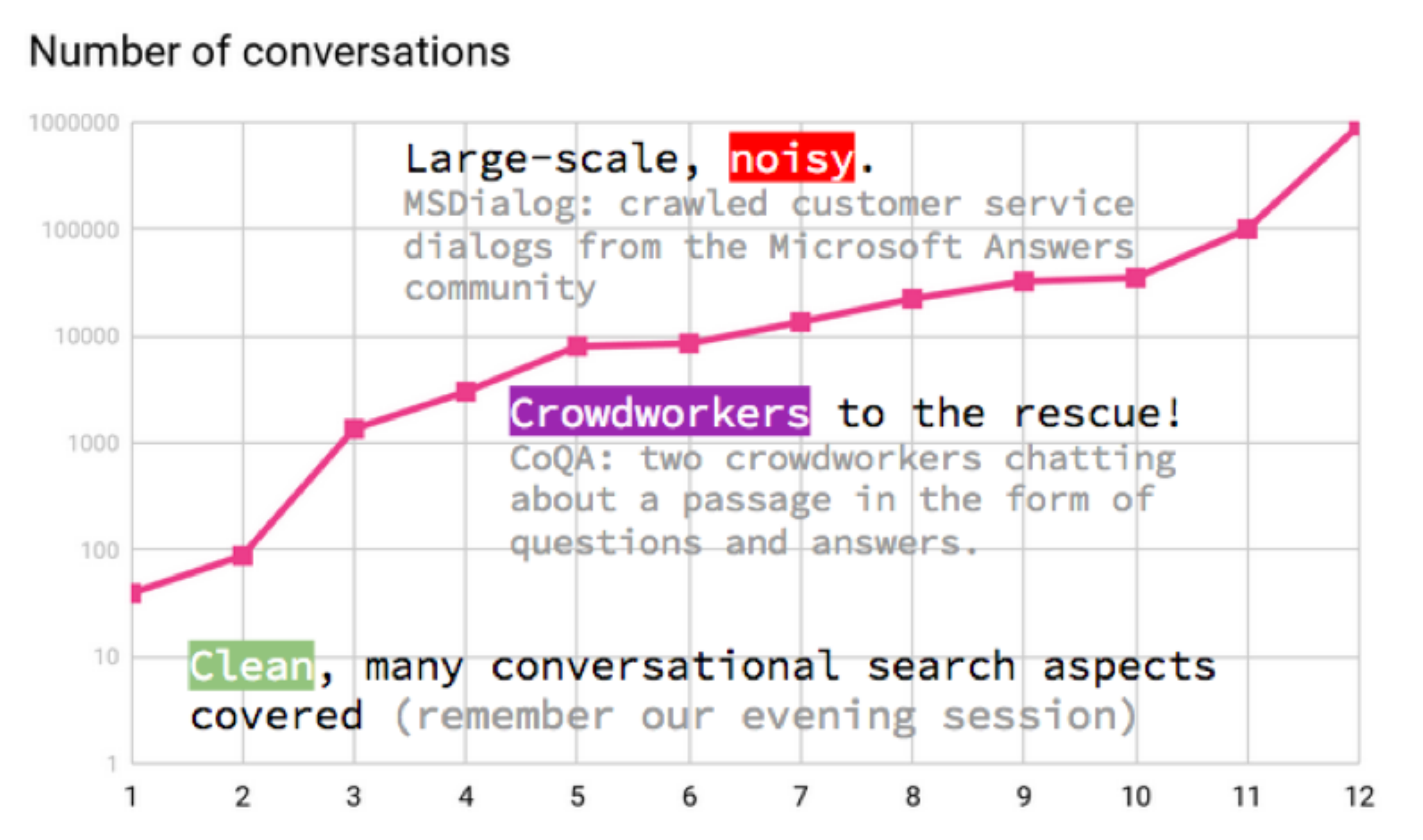}
\caption{Overview of the dataset sizes of 12 recently introduced conversational datasets that are multi-turn, non-chit-chat and human-to-human.}\label{fig:datasets}
\end{figure}

A particular challenge of conversational search datasets is to meaningfully collect and build large-scale datasets (required for neural net-based training regimes). Consider Figure~\ref{fig:datasets} where we plot the number of conversations across 12 recently introduced conversational datasets (such as MSDialog, UDC, CoQA, Frames, SCS and others). Even the largest dataset has fewer than a million conversations, while the smallest ones have fewer than 100 conversations. Importantly, the larger datasets are usually crawls of large fora (e.g. Stack Overflow or other technical fora) with little to no additional labelling to enable a range of conversational tasks. At the other end of the spectrum we have very small, but also very clean and well-annotated datasets that are very useful to analyze conversations but not sufficient to train today's machine learning algorithms.

\subsubsection{Measuring Conversational Searches and Systems}

\begin{figure}[ht]
\centering
\includegraphics[width=.6\columnwidth,angle=270]{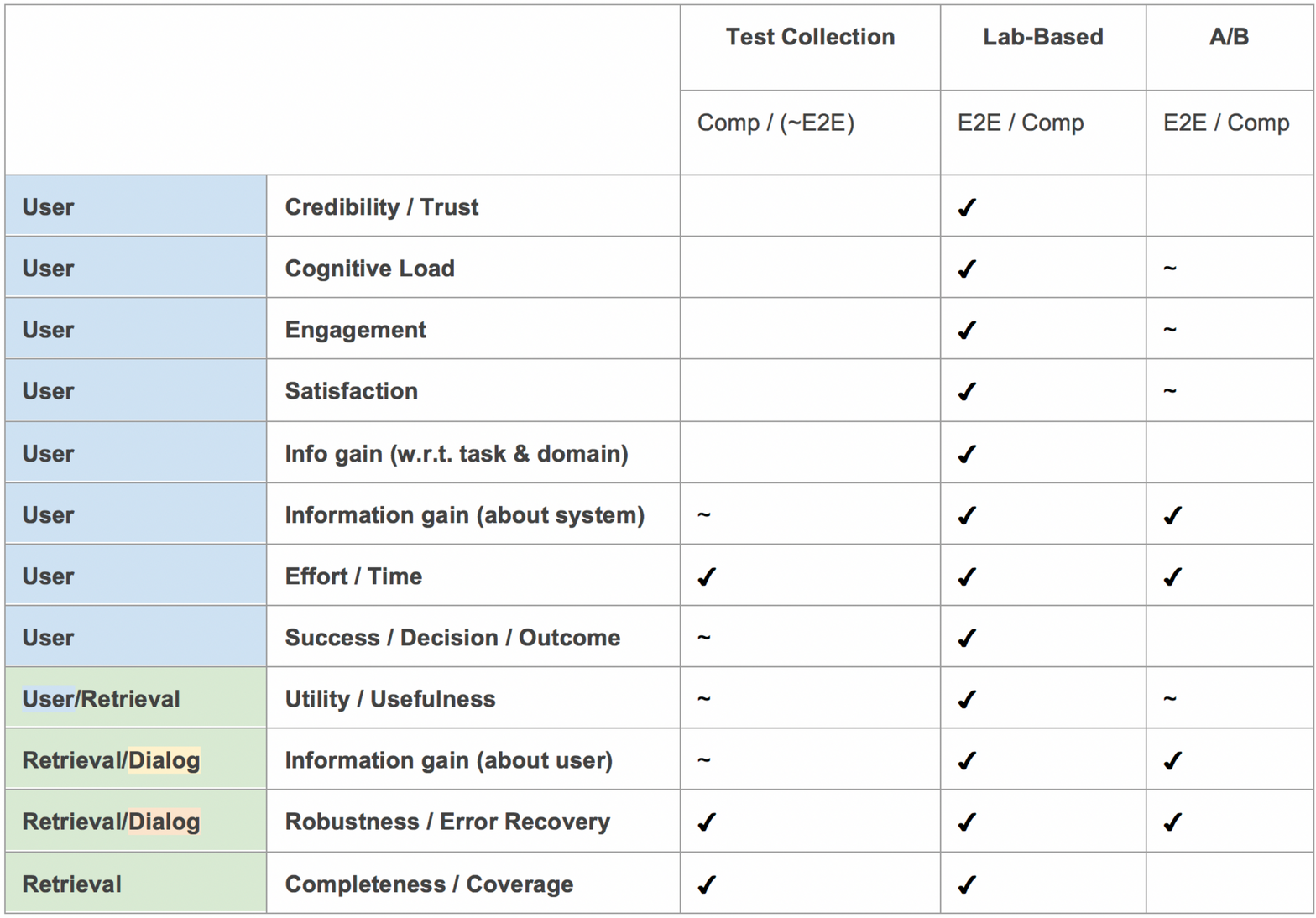}
\caption{A summary of evaluation criteria and evaluation methodologies for component-based and/or end-to-end evaluation of conversational search systems.}\label{fig:eval}
\end{figure}

In Figure~\ref{fig:eval}, we have enumerated a number of different dimensions in which we may wish to evaluate CS/CSS by, whether they are mainly user-focused, retrieval-focused or dialogue-focused. Lab-based and A/B testing will typically involve a complete (or simulated) system setup and thus facilitate end-to-end (e2e) evaluation. However, given the highly interactive nature of CS it is unlikely that a reusable test collection will be able to be developed to support any serious e2e evaluations---test collections should be able to support component level evaluation.

Ideally, the measures used should scale. That is, if the measure is used at the component level, then it should inform as to how that measure would change the e2e experience. 

Note that in the table ticks indicate that this measure can be done using test collection, lab-based or A/B testing, while ~ indicates that it might be possible or could be done via a proxy.

The different dimensions suggest that many trade-offs are likely to arise during the conversational search. For example, higher effort may be indicative of a poor CS experience, but could equally be indicative of a good conversational search experience - as it depends on how much the user gains from the experience in terms of how much they learn about the topic, the domain (and the search space) and the system (and its affordances). However, for longer term measures such as trust, it is dependent on the cumulative experiences and the successes/decisions/outcomes that result from the conversations. For example, if K buys the Nike's but finds them later for a lower price, or buys them and finds out that they are not as comfortable as described---then they may be be subsequently unhappy, and thus have less trust in the system. 

From Figure~\ref{fig:eval}, it is clear that the measures are not different from those used in interactive information retrieval - however, depending on the form of conversational search, certain dimensions are likely to be more important than others.

\abstracttitle{Evaluating Conversational Search - An Alternate Perspective}
\abstractauthor[Rishiraj Saha Roy, Avishek Anand, Jens Edlund, Norbert Fuhr, and Ujwal Gadiraju]{Rishiraj Saha Roy (MPI für Informatik - Saarbrücken, DE), Avishek Anand (Leibniz Universität Hannover, DE), Jens Edlund (KTH Royal Institute of Technology - Stockholm, SE), Norbert Fuhr (Universität Duisburg-Essen, DE), and Ujwal Gadiraju (Leibniz Universität Hannover, DE)}
\license

\subsubsection{Description}

Saracevic~\cite{Saracevic/Covi:00} describes a nice framework for the evaluation of digital libraries, which can be applied to all kinds of information systems. For setting up an evaluation, they define five aspects to be addressed. We describe each of these in the following, but reshape the original definition for our case of evaluating conversational search systems (CSS).
\begin{enumerate}
    \item \textbf{Construct} for evaluation: What parts of the system are to be evaluated? Roughly speaking, we can distinguish between \textit{content} (the underlying corpora or knowledge bases), \textit{system} (the software and hardware) and  \textit{users} with their usages (tasks they use the system for). Most evaluations will focus on certain parts of a CSS (e.g.\ retrieval method or dialog component), but often the evaluation as such cannot be performed without employing the other CSS parts,
    \item \textbf{Context} of evaluation: Here the goal, framework, viewpoint or level(s) of the evaluation have to be defined. E.g. for a user-oriented evaluation, we might focus on the experience of real users interacting with the CSS. Current evaluation tracks for conversational search, on  the other hand, take a more system-oriented view.
    \item \textbf{Criteria}: While traditional IR evaluations mostly focus on relevance, in the context of CSS we might also be interested in aspects such as information quality/gain, reliability, credibility, correctness, resilience/robustness, completeness/coverage, user engagement and satisfaction, cognitive load, and learning about the system.

    \item \textbf{Measures}: Once the criteria have been defined, appropriate measures reflecting these criteria have to be defined. Here basic principles of measurement should not be ignored: As a negative example, many current evaluations use mean reciprocal rank, which is not an interval scale and thus neither means can be computed nor parametric significance tests be applied~\cite{Fuhr:17b}. Experimental results measured this way lack any validity.
    \item \textbf{Methods}: Here we have to choose the setting in which the evaluation is to be carried out. For CSS, three major approaches seem feasible:
    \begin{itemize}
        \item \textbf{Offline experiments} are (system-oriented) evaluations without actual user involvement, where the system is given e.g. a specific request or situation from a CS dialog, and then the system's response is evaluated.
        \item \textbf{Online lab experiments} observe users interacting with a CSS, while performing a predefined task.
        \item \textbf{Living labs} are based on an operational system with users performing their everyday tasks, where the effect of certain changes to the base system is evaluated. 
    \end{itemize}
\end{enumerate}

\subsubsection{Scenarios}

Conversational search systems can support users in a diverse set of scenarios, like e.g. 

\begin{itemize}
    \item Performing an actionable task (a directed search task)
    \item Learning about a topic (an exploratory task)
    \item Occupying one's time
    \item Finding out about oneself 
    \item Making a decision
\end{itemize}

With an aim to delineate evaluation protocols in different scenarios of conversational search, we consider the following tasks: 
(a) an objective exploratory search scenario, and (b) a subjective exploration of a given topic.

\textbf{Scenario A: Objective Exploratory Search}



This scenario addresses use-cases where there is an objective specification of user needs under the closed-world assumption (in the context of knowledge management). The scenario entails recall-oriented exploratory search tasks, wherein the aspect of evaluation is the \texttt{content} and the dimension of evaluation is the entire \texttt{session}.  Evaluation in such scenarios should consider both \textit{user}, as well as \textit{system revealment} \cite{nordlie1999user,radlinski2017theoretical}. Since the user needs are objectively defined, the metric(s) used to evaluate the content that is presented and consumed pertain to information coverage. To facilitate comparison between systems the reference content that an end-user can explore is frozen. 

An Example: Consider a user with a given profile (having known attributes such as \textit{age}, \textit{gender}, \textit{medical status}) who wishes to explore information related to the topic of `diabetes'.

Content Units and Evaluation: Depending on the use-case at hand, the content units presented and consumed by the end-user through exploration can be of the order of phrases, concepts, sentences, passages or documents. For a given topic that a user wishes to explore, obtaining an expert reference collection of content units, such as a list of facts that should be retrieved, is a prerequisite for evaluation. By measuring session coverage of reference content units either manually or automatically, we can evaluate the aspect of `content'. Such evaluation can be carried out in online settings where users interact with a CSS. Note that several other user dependent criteria (as described earlier) are relevant and important to evaluate alongside the content evaluation.

\textbf{Scenario B: Subjective Exploration of Topic}
\label{sec:scenario2}

This scenario is driven by a user's need for discovering more information about a topic of interest. Such information needs are
rarely one-off: the user issues a series of follow-up queries around an entity or a theme. Just like conversing with a human,
follow-up utterances are often ad hoc, ungrammatical and incomplete with respect to entities and relations. Let's consider
the following example:

\begin{Snugshade}
$q_1$: What are some of Spielberg's recent movies?

$r_1$: Ready Player One, BFG, Bridge of Spies, The Post,....

$q_2$: And the lead actor in The Post was...?

$r_2$: Tom Hanks

$q_3$: Oh, and what are some of \textit{his} recent ones?

$r_3$: Toy Story 4, Mamma Mia 2, The Post, ...

$q_4$: By the way, who's he married to now?

$r_4$: Rita Wilson. Do you also want to know whom he was married to earlier...?

$q_5$: Just shut up.
\end{Snugshade}

The scenario is driven by a sequence of questions $q_i$ (equivalently queries or utterances) by a user over turns $i$.
At each turn, the agent provides a response $r_i$, that could be an answer (say, $r_3$) or a clarification question ($r_4$),
or a mixture of the two ($r_4$).
Since this is a user-driven exploration of topic, unlike Scenario 1, \textit{completeness} is not a concern, i.e, the user
does not want to know everything about Spielberg's movies, but only what he asks for, over the course of the conversation.
As a result, the performance over turns is key to user satisfaction (as opposed to only an end-of-session evaluation in Scenario 1).
This brings us to the observation that there are multiple evaluation criteria at work in this scenario (including but not limited to
the following):
\squishlist
    \item \textbf{Relevance} of the answer at each turn
    \item \textbf{Engagement} of the user over the conversation
    \item \textbf{Effort} that the user spends over the conversation
\squishend

Note that these dimensions are \textit{orthogonal}:
a user may leave the conversation after two turns after receiving exactly
correct answers (high relevance but low engagement+low effort), he may be highly engaged by spending a lot of time
hearing (ultimately) irrelevant passages from the agent (high engagement but low relevance+low effort), he may spend
a lot of effort in issuing many follow-up queries without ultimately locating relevant answers (high effort but
low relevance+low engagement), and so on.

\textbf{A bottom-up approach.} Having set up this initial premise, we needed to decide whether \textit{current benchmarks} are suitable for 
evaluation:
for example, the TREC CAsT collection (\url{http://www.treccast.ai/}),
CoQA~\cite{coqa:2019}, QuAC~\cite{quac:2018}, QBLink~\cite{qblink:2018},
ShARC~\cite{sharc:2018},
ConvQuestions~\cite{convex:2019}, and so on. After going through these benchmarks, we 
found that the CAsT corpus and ConvQuestions are most applicable to 
the scenario, but have one overarching limitation: they assume that
there is \textit{only one flow} of the conversation. Let us elaborate on this
a bit: in CAsT, the follow-up queries do not really depend on the
answer passages retrieved by the previous queries. In other words,
user intent in a given query can always be inferred by looking at past queries alone.
In ConvQuestions, follow-up utterances are indeed influenced by
previous answers, but are limited to questions with only a few correct
answers. In reality though, \textit{list questions} with several relevant
responses are highly common (like \textit{spielberg movies},
\textit{barcelona footballers}, and \textit{nobel peace prize winners}).

\begin{figure} [t]
	\centering
	\includegraphics[width=\textwidth]{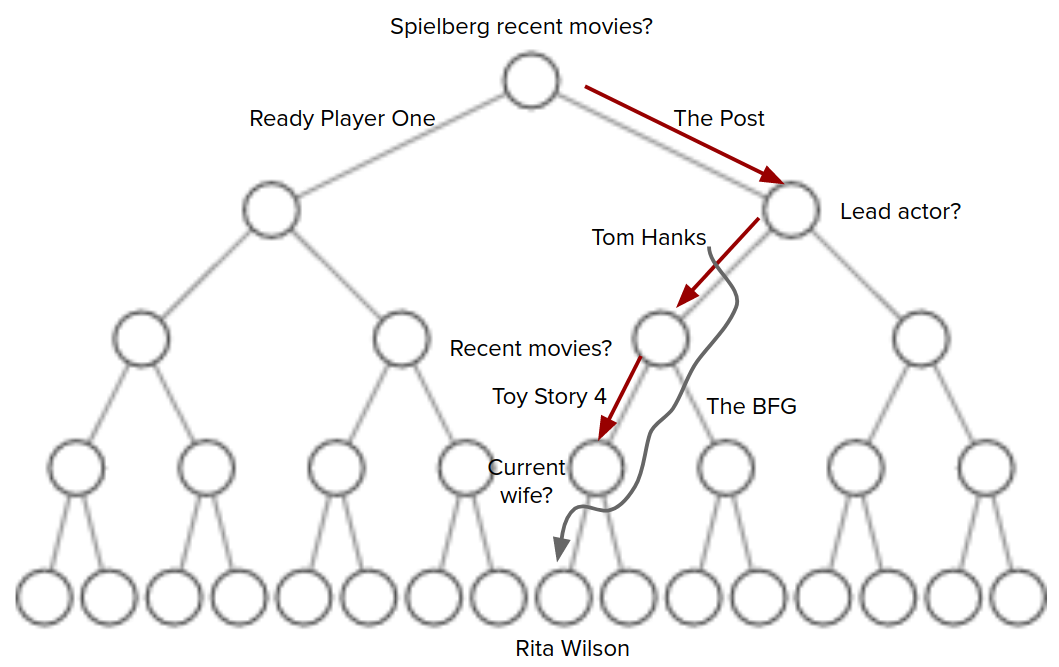}
	\caption{An illustration of the non-determinism in 
	an exploratory conversation.}
	\label{fig:flow}
\end{figure}

Even though these shortcomings appear different on surface, we posit that
there is one underlying reason that both of these evaluation testbeds shied away from a more realistic setup.
This reason is that 
having current or previous responses affect follow-up queries, or
allowing for multiple correct responses, introduces
\textit{non-deterministic flows} through the conversation tree
(Figure~\ref{fig:flow}). This
is the core problem that we identified and tried to address. 

To elaborate, when a user issues $q_1$, there may be more than
one correct $r_1$. The next query $q_2$ may be influenced by 
each of these correct $r_1$, resulting in a branching effect. 
Iteratively, $q_3$ could depend on one or more concepts from
$q_1, r_1, q_2$, and $r_2$. This is illustrated in the 
somewhat ad hoc \textit{flow tree} in Figure~\ref{fig:flow}, where \textit{nodes}
represent conversational queries and \textit{edges} mark the possible 
correct responses to the corresponding parent node.

\textbf{Proposal towards a solution.} \textit{Enumeration} of all such
possible paths (an example is marked by arrows in the figure) is
exhaustive and all paths cannot, for all practical purposes,
be included in
a static benchmark suitable for offline evaluation. A natural 
alternative is to \textit{sample} some of these paths -- that can perhaps
be collected from a few runs with independent users interacting
with one or more online systems. But the sheer volume of possible
paths and resource constraints (in terms of time, users, and money)
makes the sparsity of paths in
any such effort inevitable (only a small sample of paths can 
be realistically contained in any benchmark). Moreover, such 
a benchmark can be created only once, and should be suitable
for evaluating \textit{future systems} that have not yet been built.
Designing benchmarks and methods often result in a chicken-and-egg
paradox.

So it is possible that
a future system retrieves correct answers and passes through acceptable
paths in the tree, that are unfortunately absent in the \textit{static
snapshot} of flows. Such a situation would be unfair to the system
under test, and if each participant system makes their own incremental
update to the benchmark, such a move will also be frowned upon by
the community at large.

As a result of these observations, an initial proposal is as follows.
Some sample paths can be instantiated by any reasonable mechanism:
user studies, pooling systems, or collecting most salient responses at each step.
The system should be evaluated at each turn independently, while
keeping the history (equivalently, the path, or states) leading up to
that state fixed. To be specific, a system has to respond at 
turn $i$ with $r_i$, given a fixed $\{q_{i-1}, q_{i-2}, \ldots, q_1;
r^*_{i-1}, r^*_{i-2}, \ldots r^*_1; r_{i-1}, r{i-2}, \ldots r_1\}$. Here,
the $r^*$'s refer to gold responses and the $r$'s to the
system-generated responses.

\textbf{Shortcomings.} Our proposal has two drawbacks as of now:
\squishlist
\item \textbf{Evaluation of robustness:} To some extent, since the history building
up to a level in the tree is given upfront, the ability to evaluate
a system
to recover from past failures (resilience) is limited (since gold responses are
also included).
\item \textbf{Incorporation of clarifications:} Another open concern is the incorporation of
clarification questions, that is integral to most conversational
information access scenarios. Clarification questions are rather
flexible in their formulations, and including hard-coded verbalizations in any benchmark still appears artificial.
\squishend

\subsubsection{Broader Impact}

Current offline evaluation schemes are rather limited and unrealistic in evaluating exploratory conversational tasks.
Towards this, our envisioned  proposals could provide possible guidance in evaluating both objective and subjective exploratory tasks.
Specifically, our proposals focus on recall oriented evaluation of objective tasks and bottom up approaches exploiting non-deterministic flows for subjective evaluations.
The impact of these proposals hopefully will extend the current realm of evaluation schemes.

\subsubsection{Suggested Reading}

Some of the existing datasets and benchmarks for offline evaluation of conversational systems focus on sequential question answering on text~\cite{coqa:2019,quac:2018,sharc:2018}, knowledge graphs~\cite{csqa:2018,convex:2019} and tables~\cite{iyyer2017search}.


\abstracttitle{Modeling Conversational Search}
\abstractauthor[Elisabeth André, Nicholas J. Belkin, Phil Cohen, Arjen P. de Vries, Ronald M. Kaplan, Martin Potthast, and Johanne Trippas]{Elisabeth André (Universität Augsburg, DE), Nicholas J. Belkin (Rutgers University - New Brunswick, US), Phil Cohen (Monash University - Clayton, AU), Arjen P. de Vries (Radboud University Nijmegen, NL), Ronald M. Kaplan (Stanford University, US), Martin Potthast (Universität Leipzig, DE), and Johanne Trippas (RMIT University - Melbourne, AU)}
\license

\subsubsection{Description and Motivation}
 
An information-seeking system cannot carry out a two-way conversation to make a search more effective unless it maintains interpretable models of its own capabilities and resources, its beliefs about the goals and capabilities of the user, the history and current state of the search process, the context of the search, and other strategies and sources that might satisfy the user's information need. The reflection and self-awareness that these models support enable conversations that help the system and user come to a common understanding of the user's underlying objectives and help the user understand what the system can and cannot do. This should result in a shared plan for executing a successful search. The models are refined or reconstructed through the course of the conversational interaction, as intermediate results are presented and discussed, the search mission is clarified, and new goals and constraints come to light. Importantly, the system's strategic behavior is guided by its ability to inspect the explicit representations of intents, capabilities, and history that the evolving models encode.

In order for a conversational system to talk about a topic, it needs to have a model of that topic.  Current deeply learned systems that are trained from prior conversational interactions about arbitrary topics incorporate latent topic models. However, training such a system would require a huge amount of conversational data about that topic, an effort that would be infeasible for conversational search tasks.  Rather, a more fruitful approach may be a factored model that separately models conversation, as applied to information-seeking tasks. Thus, systems would learn how to talk separately from the specific content.

Conversational search systems should be collaborative in the sense that they attempt to satisfy the user's information seeking goals.  However, people do not often state what their motivating information-seeking goals are, and their specific information requests may not literally state what they are looking for.  The conversational search system of the future should interact collaboratively with the user to narrow down the interpretation of the user's desires, especially in the face of search failures, vague descriptions, unstructured digital information, non-digital information, and non-federated information sources, such as a museum's archives.
 
Thus, in order for a conversational system to be helpful, it needs a model of the task that motivates the information-seeking request.  Such a model would enable the conversational system to find alternative approaches to achieving the higher-level motivating goal when a failure occurs. Additionally, the conversational system would need a model of the user, especially if the information-seeking task is extended over time, in order that the system does not tell the user what it believes the user already knows. The user model should contain models of what the user knows, is intending to do or come to know, what s/he has already done, etc.  Such models could be derived from general background knowledge and from prior interactions with the system. Among the elements of the user model should be a model of what the user thinks the system can do, what it contains/knows, etc. The conversational search system will need to reveal its capabilities during interaction because it cannot display all its capabilities as menu items.  The system will also need a model of itself and models of other non-federated systems, in order that it be able to provide information that it is incapable of handling a request, but the user should inquire with another system that may contain the desired information. 
During the conversation, the user may state, or the system may request, information about the task or goal that is motivating the user's information need.  In order to understand the user's natural language response, the system will need to build its own model of the user's goals, intentions, tasks, and planned actions.   Such a model will need to be precise enough to inform the search system, but not require such precision and certainty that it cannot handle vague user responses. Indeed, part of the conversational search system's collaborative task is to gradually elicit such information and in order to narrow down such vague requests. The model of the task should at least provide parameters and actions that the information system can use to perform such sharpening.

\subsubsection{Proposed Research}

Humans have the ability to infer information about the user's beliefs and wants based on the situative and conversational context and consider this information when performing search tasks with others. For example, we might tell somebody leaving the house where to find an umbrella even when it is currently not raining, but considering that it might rain according to the weather forecast. Current search engines tend to take a macroscopic view and present the users with a number of options they might be interested in. For example, one of the authors of this abstract was provided with suggestions of hotels in cities she has visited before even though she had no intention to visit most of the cities again. While such an unsolicited collection might inspire people to explore new ideas, there are situations where users expect more selective results based on a specific search request. To accomplish this task, a system requires a deeper understanding of the user's desires, beliefs and intentions as well as the situational and conversational context. In the area of cognitive sciences, such an ability is called ``Theory of Mind''.
In many applications, such as the medical domain, it is critical to know how a system retrieved its search results, how confident it is about their sources and how results from different sources have been integrated. A system that is able to explain its behaviors is likely to increase user trust. Thus in addition to a model of the user's wants and beliefs, an explicit representation of the system's self-model is required. An explicit model of the people's and system's wants and beliefs is a necessary prerequisite for collaborative conversational search where the system, for example, asks for additional information from the user or refers to third parties to accomplish the user's initial search request.

Despite significant attempts to formalize models of the users' and the system's belief and wants for dialogue systems, this research has found surprisingly little attention in conversational search. We do not argue that all applications require deep models and explanations. In particular, users might feel overwhelmed by a system revealing too many details on its inner workings.
\begin{enumerate}
\item
Investigate how conversational search may be enhanced by a model of the users' beliefs and wants
\item
Enhance conversational search by a reflective mechanism that explains the applied search mechanism and the accessed sources
\item
Explore techniques to find a good balance between macroscopic and microscopic modeling and explanation
\end{enumerate}

\abstracttitle{Argumentation and Explanation}
\abstractauthor[Khalid Al-Khatib, Ondrej Dusek, Benno Stein, Markus Strohmaier, Idan Szpektor, and Henning Wachsmuth]{Khalid Al-Khatib (Bauhaus-Universität Weimar, DE), Ondrej Dusek (Charles University - Prague, CZ), Benno Stein (Bauhaus-Universität Weimar, DE), Markus Strohmaier (RWTH Aachen, DE), Idan Szpektor (Google Israel - Tel-Aviv, IL), and Henning Wachsmuth (Universität Paderborn, DE)}
\license

\subsubsection{Description}

Search, in a broader sense, means to satisfy an information need of a person. Conversational search, in particular, restricts the exchange of information to achieve this goal to natural language primarily (in contrast to having access to powerful display, for instance). Although a conversation may be pleasant to the information seeker, it usually implies a reduction in bandwidth: Which of the possibly many search refinement criteria should be asked first by the system? When to get what piece of information from the information seeker? Which retrieved search result should be shown first? 

A conversational search system definitely introduces a bias when choosing among questions and results, and it may frame the entire information seeking process. This raises the need for a conversational search system to explain its decisions. Even more, the conversational search system may implicitly tell the information seeker what are the important concepts related to the information need and may change the seeker's beliefs on the topic. Argumentation technology provides the means to address these and related issues.

\subsubsection{Motivation}

Argumentation and explanation are required for different purposes in conversational search. They can be essential to justify each move the system takes in the conversation, especially if the information seeker explicitly requests such information. Furthermore, argumentation is a fundamental mechanism to acknowledge different viewpoints of a discussed topic. Accordingly, argumentation technology may be used for result diversification or aspect-based search within conversational settings.

An exemplary conversational search scenario where argumentation plays a key role is scholarly research. When an information seeker attempts, e.g., to search for the best venue to submit a paper to or aims to find the most influential studies for a concrete research topic, it is highly beneficial that the system explains its answers during the conversation and even supports them with high-quality evidence.

\subsubsection{Proposed Research}

To build new computational models of argumentative conversational search, appropriate training data is required first. We propose to start with existing datasets with conversational argumentative content, such as debate portals and forum discussions (e.g., debate.org, Reddit ChangeMyView, Wikipedia talk pages, or news comments) and community question answering platforms, such as Quora \cite{omari:2016}. However, these datasets need to be filtered to focus on search scenarios only. We believe that this can be done (semi-)automatically by following the role and engagement of the seeker in the debate. Additional non-search data as well as data from wiki-like debate portals (e.g., idebate.org) can be used later to improve argumentation capabilities of the models.

To further understand the topic and to support more efficient model training, we propose developing a specific annotation scheme related to conversational search, building upon works of \cite{trippas:2017}, \cite{alkhatib:2018}, and \cite{vakulenko:2019a}. This scheme should roughly include the following layers:

\begin{itemize}
\item 
{\bf Conversational layer.} Argumentative relations, speech acts, rhetorical moves.
\item 
{\bf Demographics layer.} Socio-demographic indicators of participants as far as available, involvement of the seeker.
\item 
{\bf Topic layer.} Specific domain concepts, frames.
\end{itemize}

Furthermore, the annotation should clarify why and how each specific conversation relates to search and to a conversational need as well as why argumentation or explanation are needed to satisfy this need. As the immediate next step, we propose to run a small-scale annotation pilot study which will result in a theoretical analysis of argumentation strategies in conversational search and in data annotation guidelines tested for annotator agreement.

\subsubsection{Research Challenges}

When providing information within the conversation between a system and an information seeker, the system needs to incrementally decide upon three basic questions matching concepts from research on rhetoric and argumentation synthesis \cite{wachsmuth:2018}:

\begin{enumerate}
\item 
{\bf Selection.} How to select information, i.e., what to convey to the seeker?
\item 
{\bf Arrangement.} How to arrange the information, i.e., what to say first and what later?
\item 
{\bf Phrasing.} How to phrase the information, i.e., what linguistic style to use?
\end{enumerate}

A question arising specifically in argumentative contexts is whether the way the system provides the information should be personalized towards the profile of a specific seeker or should stay general to all seekers. A related issue is the possibility and extent of learning from user-provided information and user feedback. Also, there is a trade-off between the conciseness and the comprehensiveness of the arguments and explanations given for certain information or for the behavior of the system. 

As indicated above, however, the most immediate challenge is that no corpora are available so far that sufficiently allow carrying out the research that we propose. We therefore argue that the first challenges to be tackled are the following:

\begin{itemize}
\item 
{\bf Data.} The acquisition of a corpus for studying argumentation in conversational search. 
\item 
{\bf Annotation.} The annotation of the corpus towards the scheme outlined above.
\end{itemize}

\subsubsection{Broader Impact}

Integrating argumentation and explanation in conversational search will help elevate the retrieval of information from providing documents in a search interface to providing contextual information about sources, viewpoints, potential biases, and conventions in a more natural and dialogue-oriented way. Having explicit structures for argumentation and explanation in search allows information seekers to ask clarification and justification questions. Also, it can help the seekers to build better mental models of the underlying information retrieval processes. This will also enable to navigate different perspectives of controversial debates and thereby has the potential to overcome some of the pressing challenges of search today including filter bubbles, bias in information provision, or misinformation.


\abstracttitle{Scenarios that Invite Conversational Search}
\abstractauthor[Lawrence Cavedon, Bernd Fröhlich, Hideo Joho, Ruihua Song, Jaime Teevan, Johanne Trippas, and Emine Yilmaz]{Lawrence Cavedon (RMIT University - Melbourne, AU), Bernd Fröhlich (Bauhaus-Universität Weimar, DE), Hideo Joho (University of Tsukuba - Ibaraki, JP), Ruihua Song (Microsoft XiaoIce- Beijing, CN), Jaime Teevan (Microsoft Corporation - Redmond, US), Johanne Trippas (RMIT University - Melbourne, AU), and Emine Yilmaz (University College London, GB)}
\license

Our working group identified scenarios that invite conversational search. What emerged is (1) no other modality available (or best modality is different), (2) the task invites conversation. In this document, we motivate these key scenarios and propose research around prototypical tasks in this space. The associated key research challenges were identified in collecting, constructing and representing the rich multimodal contextual information of conversational search, summarizing and presenting the results in speech-only scenarios, design of conversational strategies and in evaluating the dialogue and search systems. Collaborative conversational search adds further challenges that consider the potentially highly interactive, multimodal and synchronous communication between humans and agents.

\subsubsection{Motivation}

Natural language conversation is not always the best way for a person to search. Conversational search makes the most sense when (1) the situation requires that a person uses an interaction modality that is better suited to conversational interaction than conventional input and output methods, or (2) when the task requires significant context and interaction. In this section we expand on scenarios related to these two cases, and also explore when conversational search might not be the right approach.

\paragraph*{Interaction and Device Modalities that Invite Conversational Search}

Conversational search is particularly useful when a person's search interactions will be via a modality other than the traditional screen, keyboard, and mouse. This may be because people do not have immediate access to a conventional computer (e.g., they are driving or cooking), are unable to use one (e.g., due to impaired vision or literacy constraints) or they might be simply not very proficient in typing. It may also be because other form factors that are more readily available that lend themselves to conversation e.g. a smartwatch. Furthermore, many modern form factors, like smart speakers, earbuds, or AR/VR systems, have no keyboard and are designed around speech in- and output. Because speech lends itself to far-field interaction, it enables a person to search without actually going to the device and makes it easy for multiple people to simultaneously interact with the system.

\paragraph*{Tasks that Invite Conversational Search}

Search tasks currently supported by non-conventional modalities tend to be simple and fact-finding in nature (e.g., ``Cortana, what is the weather in Frankfurt?''). However, we expect these systems starting to address more complex tasks (i.e., tasks where different information units need to be inspected and compared) as conversational search capabilities improve. Furthermore, conversation is good for building shared context and common ground, and tasks that require much contextual information – on the part of one or more searchers, the system, or shared between them – invite conversational search even when someone is using conventional modalities.

For this reason, conversational search is likely to be particularly useful for exploratory search tasks where the searcher wants to learn about an area. Such tasks typically require clarification of the searcher's need, and the search process may be so complex that it needs to be decomposed into pieces. Conversation can help guide this process while maintaining the larger picture. Conversational search can also be useful where sense-making is required to understand the content the system provides. In contrast to exploratory search, with casual information seeking the searcher does not have a particular goal and just wants to be entertained in a similar way as when browsing a news feed. As an example, a news article might serve as a starting point which sparks interest in further information about some mentioned facts which could be verbally expressed without the need of going to a search engine. In such scenarios, users are often looking to cognitively and affectively make sense of how the world works and why or they might want to relate some provided information to their personal environment and life. Conversational search may also be useful when a balanced view is important to understand a particular issue and come up with solutions to the issue.

Finally, conversational search makes much sense in contexts where multiple people are involved and there is a shared context. People communicate with each other via conversation, in meetings, via email and text chat, and even through things like comments in documents. A conversational search system is likely to be a good way to address information needs that come up in the course of these conversations, and conversational search tasks seem particularly likely to be collaborative.

\paragraph*{Scenarios that Might \textit{not} Invite Conversational Search}

Conversational search is not always a good idea and can add overhead for simple information needs where existing channels already work well. Conversations carry cognitive load and offer limited bandwidth. The traditional keyword search paradigm thus probably makes more sense than conversation when a person's modality is not constrained, it is easy for them to describe their information need via querying, and the task requires high bandwidth output that is well served by a ranked list. This may be particularly true for highly ambiguous situations where quick iteration is useful, as people often have a hard time understanding the limits of conversational systems, and recovering from failure in natural language can be hard. Speech based systems can also be problematic in social situations where they can disrupt others or unintentionally expose private information.

\subsubsection{Proposed Research}

We propose that conversational search research focus on addressing these modalities and tasks. Prototypical scenarios that look at interaction and modalities that invite conversational search often include speech, and must handle noise, address distraction and errors, and be aware of social context. Some examples include:

\begin{itemize}
\item
Mechanic fixing a machine, wants to know something to help them do a better job.
\item
Two people searching for a place to eat dinner via speech while driving. The system asks for their preferences and mediates their discussion of the options.
\end{itemize}

Prototypical scenarios that address tasks that invite conversational search are ones that require significant exploration, interaction, and clarification. Examples include:
\begin{itemize}
\item
Learning about a recent medical diagnosis. Includes the person asking for general information, the system asking clarifying questions and providing some context, and then dealing with follow up questions from the person.
\item
Following up on a news article to learn more about the topic and get additional closely or loosely related facts.
\end{itemize}

\subsubsection{Research Challenges and Opportunities}

Various research questions arise due to the multimodal aspect of conversational search, as well as due to the importance of considering the context for conversational search. Some issues particularly important in speech-based conversational systems in general also apply to conversational search such as the personality of the system as well as privacy and security issues which we do not discuss here.

\paragraph*{Context in Conversational Search}

With the multimodality and richer scenarios for conversational search in mind, a variety of contextual aspects need to considered including task context, personal context (affect, cognitive load, etc.), spatial context (location, environment), or social context. General research questions regarding the context in conversational search might include: What are the contextual factors where conversational search systems are reliable to collect and process and what are not? What are effective mechanisms and models for collecting, constructing this contextual information? Are (personal) knowledge graphs and knowledge bases sufficient for representing this information? How could the system incorporate these additional sources of information into the search process?

\paragraph*{Result presentation}

Speech-only communication is a not an uncommon modality for conversational systems, and this raises specific challenges in the case of output from Conversational Search Systems, which can provide information-rich output that may be difficult to process by human consumers, due to cognitive and memory limitations. The temporally-linear and ephemeral nature of speech also limits the ability to ``scan'' results: strategies for overcoming such limitations needs to be devised, possibly including: 
\begin{itemize}
\item
Designing methods to present result summaries, or of result categories, to facilitate discussion and clarification of results of specific interest;
\item
Designing techniques to facilitate ``tagging'' of results for later reference;
\item
Designing techniques to highlight specific aspects of results to indicate their relevance. 
\end{itemize}{}

\paragraph*{Conversational strategies and dialogue}

New conversational strategies that support information seeking behaviours need to be designed: The conversational structure implemented by a system should mirror and/or support information seeking behaviour, which raises various questions such as:
\begin{itemize}
\item
How to detect and model information seeking behaviours that should be supported?
\item
What do the corresponding conversational structures/operations look like: e.g., what conversational operations support identifying the user's uncompromised information need?
\end{itemize}

Conversational search can provide opportunities to ask users clarifying questions to obtain more information about their search task, work tasks and personal condition (e.g. medical condition) for a better understanding of the users' needs, to personalise the responses to an individual user or to recover from errors. What is the structure of clarifying questions that help better understand end-users search tasks and work tasks? What are effective mechanisms for constructing such clarification questions? What level of personification is desirable in conversational search tasks?

\paragraph*{Evaluation}

Availability of different modalities would also require the design of new evaluation methodologies for conversational search which should consider implicit and explicit satisfaction signals present in responses from users including affect, tone of voice and cognitive load. In a dialogue we can also explicitly ask for feedback or implicitly provoke conversational responses that inform the evaluation.

\paragraph*{Collaborative Conversational Search}

Person-to-person communication scenarios are a particularly promising application field of speech-based conversational search since the need for search might naturally emerge from a conversation. Here, the general challenge is to augment unobtrusively a potentially highly interactive, multimodal and synchronous communication of humans being co-located or at different locations (e.g., Skype). Conversational agents need to be aware of the roles of the users and social context of the communication. Furthermore, when multiple people are involved, conflicts, different points of view and different goals and interests are an inherent part of the conversational search process.

Particular research challenges for collaborative scenarios include the identification of prototypical, collaborative information seeking processes, the extraction of an information need from a conversation happening between people and the construction of a corresponding representation of the information seeking task. Work on research questions such as how personal knowledge graphs of individual users can be merged into a group knowledge graph or how to design effective multi-party NLP systems can provide the necessary building blocks for collaborative conversational search systems.


\abstracttitle{Conversational Search for Learning Technologies}
\abstractauthor[Sharon Oviatt and Laure Soulier]{Sharon Oviatt (Monash University - Clayton, AU) and Laure Soulier (UPMC - Paris, FR)}
\license

Conversational search is based on a user-system cooperation with the objective to solve an information-seeking task. In this report, we discuss the implication of such cooperation with the learning perspective from both user and system side. We also focus on the stimulation of learning through a key component of conversational search, namely the multimodality of communication way, and discuss the implication in terms of information retrieval. We end with a research road map describing promising research directions and perspectives.

\subsubsection{Context and background}

\paragraph*{What is Learning?}

Arguably, the most important scenario for search technology is lifelong learning and education, both for students and all citizens. Human learning is a complex multidimensional activity, which includes procedural learning (e.g., activity patterns associated with cooking, sports) and knowledge-based learning (e.g., mathematics, genetics). It also includes different levels of learning, such as the ability to solve an individual math problem correctly. It also includes the development of meta-cognitive self-regulatory abilities, such as recognizing the type of problem being solved and whether one is in an error state. These latter types of awareness enable correctly regulating one's approach to solving a problem, and recognizing when one is off track by repairing momentary errors as needed. Later stages of learning enable the generalization of learned skills or information from one context or domain to others-- such as applying math problem solving to calculations in the wild (e.g., calculation of garden space, engineering calculations required for a structurally sound building).

\paragraph*{Human versus System Learning}

When people engage an IR system, they search for many reasons. In the process they learn a variety of things about search strategies, the location of information, and the topic about which they are searching. Search technologies also learn from and adapt to the user, their situation, their state of knowledge, and other aspects of the learning context \cite{collinsthompson2017}. Beyond adaptation, the engagement of the system impacts the search effectiveness: its pro-activity is required to anticipate user's need, topic drift, and lower the cognitive load of users \cite{Tang2019}. For example, when someone is using a keyboard-based IR system of today, educational technologies can adapt to the person's prior history of solving a problem correctly or not, for example by presenting a harder problem next if the last problem was solved correctly, or presenting an easier problem if it was solved incorrectly.

Based on conversational speech IR systems, it is now possible for a system to process a person's acoustic-prosodic and linguistic input jointly, and on that basis a system can adapt to the person's momentary state of cognitive load. The ideal state for engaging in new learning would be a moderate state of load, whereas detection of very high cognitive load might suggest that the person could benefit from taking a break for some period of time or address easier subtopics to decomplexify the search task \cite{Awadallah2014}.

\subsubsection{Motivation}

\paragraph*{How is Learning Stimulated?}

Based on the cognitive science and learning sciences literature, it is well known that human thought is spatialized. Even when we engage in problem-solving about temporal information, we spatialize it \cite{JohnsonLaird1996}. Since conversational speech is not a spatial modality, it is advantages to combine it with at least one other spatial modality. For example, digital pen input permits handwriting diagrams and symbols that convey spatial location and relations among objects. Further, a permanent ink trace remains, which the user can think about. Tangible input like touching and manipulating objects in a virtual world also supports conveying 3D spatial information, which is especially beneficial for procedural learning (e.g., learning to drive in a simulator). Since learning is embodied and enhanced by a person's physical activity, touch, manipulation, and handwriting can spatialize information and result in a higher level of interactivity, producing more durable and generalizable learning. When combined with conversational input for social exchange with other people, such input supports richer multimodal input.

Based on the information-seeking point of view, the understanding of users' information need is crucial to maintain their attention and improve their satisfaction. As of now, the understanding of information need has been evaluated using relevant documents, but it implies a more complex process dealing with information need elicitation due to its formulation in natural language \cite{Aissa2018} and information synthesis \cite{Marchionini2006,White2009}. There is, therefore, a crucial need to build information retrieval systems integrating human goals.

\paragraph*{How Can We Benefit from Multimodal IR?}

Multimodality is the preferred direction for extending conversational IR systems to provide future support for human learning. A new body of research has established that when a person can use multimodal input to engage a system, all types of thinking and reasoning are facilitated, including (1) convergent problem solving (e.g., whether a math problem is solved correctly); (2) divergent ideation (e.g., fluency of appropriate ideas when generating science hypotheses); and (3) accuracy of inferential reasoning (e.g., whether correct inferences about information are concluded or the information is overgeneralized) \cite{Oviatt2013}. It is well recognized within education that interaction with multimodal/multimedia information supports improved learning. It also is well recognized that this richer form of information enables accessibility for a wider range of diverse students (e.g., blind and hearing impaired, lower-performing, non-native speakers) \cite{Oviatt2013}.

For these and related reasons, the long-term direction of IR technologies would benefit by transitioning from conversational to multimodal systems that can substantially improve both the depth and accessibility of educational technologies. With respect to system adaptivity, when a person interacts multimodally with an IR system, the system now can collect richer contextual information about his or her level of domain expertise \cite{Oviatt2018}. When the system detects that the person is a novice in math, for example, it can adapt by presenting information in a conceptually simpler form and with fewer technical terms. In contrast, when a person is detected to be an expert, the system can adapt by upshifting to present more advanced concepts using domain-specific terminology and greater technical detail. This level of IR system adaptivity permits targeting information delivery more appropriately to a given person, which improves the likelihood that he or she will comprehend, reuse, and generalize the information in important ways. The more basic forms of system adaptivity are maintained, but also substantially expanded by the integration of more deeply human-centered models of the person and their existing knowledge of a particular content domain.

Apart from the greater sophistication of user modeling and improved system adaptivity, multimodal IR systems would benefit significantly by becoming more robust and reliable at interpreting a person’s queries to the system, compared with a speech-only conversational system \cite{Oviatt2015}. This is because fusing two or more information sources reduces recognition errors. There are both human-centered and system-centered reasons why recognition errors can be reduced or eliminated when a person interacts with a multimodal system. First, humans will formulate queries to the IR system using whichever modality they believe is least error-prone, which prevents errors. For example, they may speak a query, but switch to writing when conveying surnames or financial information involving digits. In addition, when they encounter a system error after speaking input, they can switch to another modality like writing information or even spelling a word--which leads to recovering from the error more quickly. When using a speech-only system, instead the person must re-speak information, which typically causes them to hyperarticulate. Since hyperarticulate speech departs farther from the system's original speech training model, the result is that system errors typically increase rather than resolving successfully \cite{Oviatt2015}.

\paragraph*{How can user learning and system learning function cooperatively in a multimodal IR framework?}

\begin{figure}
\centering
\includegraphics[width=0.8\textwidth]{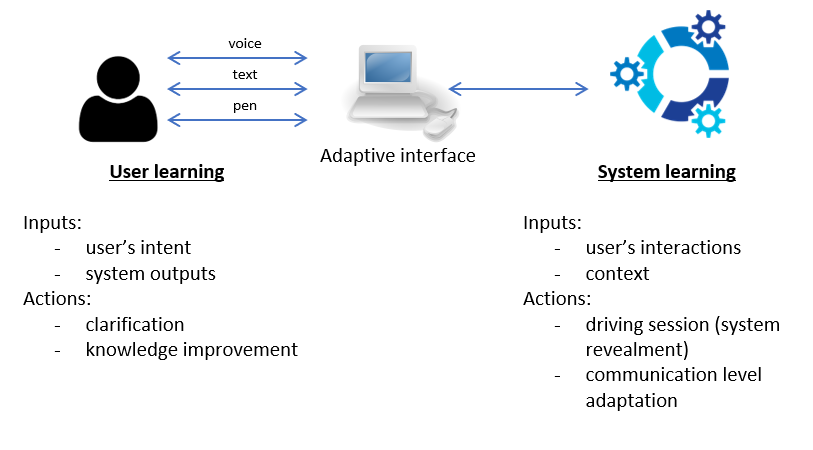}
\caption{User Learning and System Learning in Conversational Search.}
\label{fig}
\end{figure}

Conversational search needs to be supported by multimodal devices and algorithmic systems trading off search effectiveness and users' satisfaction \cite{Tang2019}. Figure \ref{fig} illustrates how the user, the system, and the multimodal interface might cooperate. The conversation is initiated by users who formulate their information need through a modality (voice, text, pen, etc). The system is expected to be proactive by fostering both (1) user revealment by eliciting the information need and (2) system revealment by suggesting what actions are available at the current state of the session \cite{Azzopardi2018a}. In response, users are able to clarify their need and the span of the search session, providing them a deeper knowledge with respect to their information need. The relevant features impacting both users and system's actions include (1) users' intent, (2) users' interactions, (3) system outputs, and (4) the context of the session (communication modality, spatial and temporal information, etc.). Several advantages of the user and system cooperation might be noticed. First, based on past interactions, the system is able to learn from right and wrong past actions. It is, therefore, more willing to target IR pieces of information that might be relevant to users. This straightforward allows reducing interactions between users and systems and lower the cognitive effort of users. Second, users being driven by increasing their knowledge acquisition experience, the system should be able to learn users' satisfaction and therefore bolster new information in the retrieval process. Altogether, these advantages advocate for a more sophisticated and a deeper user modeling regarding both knowledge and retrieval satisfaction.

\subsubsection{Research Directions and Perspectives}

\textbf{Proposed Research and Challenges: Directions for the Community and Future PhD Topics.}
Among the key research directions and challenges to be addressed in the next 5-10 years in order to advance conversational search as a more capable learning technology are the following:
\begin{itemize}
\item
Transforming existing IR knowledge graphs into richer multi-dimensional ones that currently are used in multimodal analytic research — which supports integrating information from multiple modalities (e.g., speech, writing, touch, gaze, gesturing) and multiple levels of analyzing them (e.g., signals, activity patterns, representations).
\item
Integration of multimodal input and multimedia output processing with existing IR techniques
\item
Integration of more sophisticated user modeling with existing IR techniques, in particular ones that enable identifying the user's current expertise level in the content domain that is the focus of their search and leveraging the span of the search session.
\item
Conversely, integrating analytics that enable the user to identify the authoritativeness of an information source (e.g., its level of expertise, its credibility or intent to deceive).
\item
Development of more advanced multimodal machine learning methods that go beyond audio-visual information processing and search.
Development of more advanced machine learning methods for extracting and representing multimodal user behavioral models.
\end{itemize}

\textbf{Broader Impact.}
The research roadmap outlined above would result in major and consequential advances, including in the following areas:
\begin{itemize}
\item
More successful IR system adaptivity for targeting user search goals.
\item
IR systems that function well based on fewer and briefer interactions between user and system.
\item
IR system that are more reliable and robust at processing user queries.
Expansion of the accessibility of IR technology to a broader population.
\item
Improved focus of IR technology on end-user goals and values, rather than commercial for-profit aims.
\item
Improvement of powerful machine learning methods for processing richer multimodal information and achieving more deeply human-centered models.
\item
Acceleration of the positive impact of lifelong learning technologies on human thinking, reasoning, and deep learning.
\end{itemize}

\textbf{Obstacles and Risks.}
\begin{itemize}
\item
Establishing and integrating more deeply human-centered multimodal behavioral models to advance IR technologies risks privacy intrusions that must be addressed in advance.
\item
Establishing successful multidisciplinary teamwork among IR, user modeling, multimodal systems, machine learning, and learning sciences experts will need to be cultivated and maintained over a lengthy period of time.
\item
Mutually adaptive systems risk unpredictability and instability of performance, and must be studied to achieve ideal functioning.
\item
New evaluation metrics will be required that substantially expand those used by IR system developers today.
\end{itemize}

\subsubsection*{Acknowledgements}

We would like to thank the Schloss Dagstuhl and the seminar organizers Avishek Anand, Lawrence Cavedon, Hideo Joho, Mark Sanderson, and Benno Stein for this week of research introspection and networking. We also thank the ANR project SESAMS (Projet-ANR-18-CE23-0001) which supports Laure Soulier's work on this topic.


\abstracttitle{Common Conversational Community Prototype: Scholarly Conversational Assistant}
\abstractauthor[Krisztian Balog, Lucie Flekova, Matthias Hagen, Rosie Jones, Martin Potthast, Filip Radlinski, Mark Sanderson, Svitlana Vakulenko, and Hamed Zamani]{Krisztian Balog (University of Stavanger, NOR), Lucie Flekova (Technische Universit{\"a}t Darmstadt, DE), Matthias Hagen (Martin-Luther-Universit{\"a}t Halle-Wittenberg, DE), Rosie Jones (Spotify, US), Martin Potthast (Leipzig University, DE), Filip Radlinski (Google, UK), Mark Sanderson (RMIT University, AUS), Svitlana Vakulenko (University of Amsterdam, NL), and Hamed Zamani (Microsoft, US)}
\license

\subsubsection{Description}

This working group discussed the potential for creating academic resources (tools, data, and evaluation approaches) to support research in conversational search, by focusing on realistic information needs and conversational interactions.
Specifically, we propose to develop and operate a prototype conversational search system for scholarly activities.
This Scholarly Conversational Assistant would serve as a useful tool, a means to create datasets, and a platform for running evaluation challenges by groups across the community.

\subsubsection{Motivation}

Conversational search is a newly emerging research area that aims to provide access to digitally stored information by means of a conversational user interface, that is, a dialogue-based interaction inspired and informed by human communication processes~\cite{DBLP:conf/cui/2019,phdthesis_J,phdthesis_S}. The major goal of a conversational search system is to effectively retrieve relevant answers to a wide range of questions expressed in natural language, with rich user-system dialogue as a crucial component for understanding the question and refining the answers~\cite{Allan:2012:FCO:2215676.2215678}. The respective dialogue comprises of a sequence of exchanges between one or more users and a conversational search system, which can enable multi-step task completion and recommendation~\cite{Culpepper:2018:RFI:3274784.3274788}. Several theoretical frameworks that further specify various components and requirements for an effective conversational search system have recently been proposed~\cite{Radlinski:2017:TFC:3020165.3020183,strathprints64619,Trippas:2017:PIC:3020165.3022144,DBLP:conf/ecir/VakulenkoRCR19,trippas2020towardsA}.

It is commonly recognized that only few natural conversational search corpora exist. Rather, corpora are often created through imagined needs (often in task-oriented Wizard-of-Oz studies), are inspired by logs, or come from crawls of community fora. This leads to significant research effort being planned around existing biased data and metrics, rather than data and metrics being constructed to support the most impactful research. While there have been instances of the research community interaction enabling research, such as at ECIR 2019,\footnote{\url{http://ecir2019.org/sociopatterns/}} this is relatively rare. One of our key motivations is to produce a system and corpus that contains and supports real user needs. 

Simultaneously, our community has common unsatisfied needs that appear very well suited to conversational search. Some common tasks are performed by researchers repeatedly without providing any community research value in terms of data and feedback collection, despite being relevant to many published experiments. Examples of these tasks include PC selection or finding interest profiles in EasyChair, or identifying the most relevant sessions in the Whova conference app. The collective time spent (arguably inefficiently) by our community on such tasks may far surpass the cost of creating a system that also supports research progress while providing this \emph{community value}.

\subsubsection{Proposed Research}

We propose to develop and operate a prototype conversational search system (Scholarly Conversational Assistant) that would serve as 
\begin{itemize}
\item 
a useful search tool, 
\item 
a means to create datasets for further academic research, 
\item 
and a platform for running evaluation challenges by groups across the community.
\end{itemize}

In particular, the Scholarly Conversational Assistant would allow our research community to perform a range of research-related activities. In extensive discussions, we settled on this domain for a number of reasons: (1) The data that is involved (such as papers authored, conferences/talks attended, PC memberships) is generally considered less private. Indeed most such data is already public albeit difficult to search. (2) The system is one that the members of our community would be using ourselves, giving an active knowledgeable participant base, who could contribute improvements and publish papers based on interactions observed. (3) It caters to a broad range of information needs (see below) that are currently not supported well by existing systems. (4) The relevant research groups could avoid competing with commercial providers.

A number of other possible domains were discussed, including movies, music, news, and podcasts. They have a significantly larger potential audience, yet potentially compete with commercial providers. In determining our plan, it became clear that some participants also consider interests in these areas to be highly sensitive or personal. As a critical constraint, privacy of relevant data is key (having impacted, for example, the Living Labs research \cite{Hopfgartner:2019:CEL} despite significant effort).

\subsubsection{Research Challenges}

The aim of the Scholarly Conversational Assistant system would be to enable a wide variety of research in conversational search by covering example information needs like:
\begin{itemize}
\item 
``What should I read?''---Find research on a new area of interest.
\item 
``Help me plan my attendance''---Plan what sessions to attend and whom to talk to at a conference. (Conference organizers could also use that information for optimizing room allocations.)
\item 
``Whom should I invite?''---Find conference PC, SPC, session chairs, invite speakers, etc.
\end{itemize}

Importantly, the system would log all interactions such that classes of information needs that have potential for study may be identified over time. People may evaluate the system by filling out a questionnaire, with the option of free text feedback, after each conversation (and possibly leave comments behind for individual system utterances).

\paragraph*{Connection to Knowledge Graphs}

The system would operate on a \emph{personal research graph} (PKG)~\cite{Balog:2019:PKG}, more specifically, the portion of the PKG that the user wants to share with the system. The PKG could include, among other information:
\begin{itemize}
\item 
Authorship information (which may be connected to a public citation graph),
\item 
Conference committee membership, awards, etc.,
\item 
Talks given anywhere public,
\item 
Attendance of conferences, sessions, etc.,
\item 
(in the private part) Annotations of papers, notes on talks, etc.
\end{itemize}

\paragraph*{First Steps}

The project is ambitious, but we think it can be grown incrementally:
\begin{itemize}
\item 
A starting point would be to get one ore more graduate students to start coding a tool and check it in to GitHub. It is likely that students will be able to build on top of existing infrastructure. In order for this to work, it will be necessary for a research team to own the decisions who (believes they will) get value out of such work. With a prototype system in place, one could establish a shared task at a workshop or conduct a lab study at scale. One might also design a challenge at TREC/CLEF to make use of the skeleton.
\item 
One might alternatively start by collecting evidence that such a system is something the community actually wants. Here, a sample of dialogues or information needs (that one might want to support) could be gathered.
\end{itemize}

\subsubsection{Broader Impact}

The organization of shared tasks has a long tradition in information retrieval as well as natural language processing and the dialogue community within it. In conversational search, these two communities will collaborate to build search systems that have a natural language interface as well as conversational capabilities. The breadth of potential tasks that are due to this confluence of research fields---as also identified in Dagstuhl Seminar 19461---is large. As such, developing common infrastructure and shared tasks would have high value for the community.

In particular, the outcome of shared tasks are typically large corpora and performance measures that, together, form reusable benchmarks. For example, the Cranfield-style evaluation frameworks that were adapted by TREC, or the corpora developed for the CoNLL shared tasks have had a broad impact on their respective communities at large. We expect that a conversational search challenge, too, will help to align and shape the community. 
Moreover, by developing specific shared tasks in the form of living labs~\cite{Hopfgartner:2018:ECS,Hopfgartner:2019:CEL}, we see the opportunity to apply early conversational search systems in practice as soon as possible. Here, the application domain of scholarly search, while allowing for a wide range of basic and advanced evaluation setups, may ideally transfer directly into new prototypes to enhance research itself, for instance, impacting the productivity of managing one's personal conferences schedules.

\subsubsection{Obstacles and Risks}

A variety of systems for storing and accessing research publications, reviews and conference attendance already exist. For the Scholarly Conversational Assistant to be successful, it must either be more useful than these, or potentially integrate with them. Some of the existing systems include: dblp, semantic scholar, ACM library, Google scholar, ACL anthology, open review, arXiv, Athena conference chatbot, Citeseer, Arnetminer, and arXivDigest (more on these in related reading). 

\noindent
Risks involved in operationalizing our envisaged conversational search system include:
\begin{itemize}
\item 
\emph{Privacy and data retention rules.} Ideally, the Scholarly Conversational Assistant would allow the logging of user interactions including voice input. For all personal data, the system would require a process for data access, retention and deletion as well as logging, in compliance with local regulations. Even the use of third-party speech recognizers may be sensitive depending on the location of data storage. 
\item 
\emph{Opinions != facts in indexing.} Some information that could be collected is likely to be expressed opinions rather than facts (e.g., tweets about papers). Thus, we may want to allow verification of such information before use for search and recommendation, or present it in a separate clearly-marked format with the potential for correction or deletion. Others may wish to combine private information (such as a user's personal opinions about papers), without this information being propagated. 
\item 
\emph{Speech recognition.} The use of third-party speech recognizers may be sensitive depending on the location of data storage. In addition, in the Scholarly Conversational Assistant case, the corpus contains many proper names and technical terms. A speech recognizer may require a custom language model integrating this corpus to perform well.
\item 
\emph{Personal Knowledge Graph implementation.} We would need a design that allows both cloud- and client-side storage of personal data. We need to make sure that private parts of the PKG remain private and also that users have full control over what is stored in their PKG. In case an offline dataset is created and shared, there needs to be an agreement in place that ensures that personal data would need to be removed upon request. (It should be noted that there is no way to enforce this, and ``unauthorized'' access may only be spotted if people publish using that data.)
\item 
\emph{Usage volume.} Low user participation is a concern. Beyond ensuring that the system is useful, other ways to mitigate this could include rewarding (paying) users or incentivizing them through gamification (e.g., at conferences to use the system).
\item 
\emph{Implementation.} The underlying system would require a significant effort to implement. As this would likely be contributions from different practitioners at various stages in their careers over an extended time, the contributors would naturally change. To alleviate some associated risk, a strong modularization would be beneficial, with clear interfaces and documentation. Moreover, the design of the initial prototype should be as simple as possible, with agreement of how the system's continued development is ensured during operation. The live service would also need coordination, for example, of how live experiments are planned and executed.
\item 
\emph{Operation.} Past academic systems have often been deployed on individual servers without redundancy, and potentially lacking resources for scalability. This project would likely wish to consider for this project to identify possible sponsorship from a cloud provider or host institution with significant cluster resources. The hosting decision should likely take into account long-term commitment.
\item 
\emph{Stability and reproducibility.} If used for online challenges where participants submit code that runs live, this would need to be of suitable quality to be widely used. Care would need to be taken in designing common APIs that minimize the risks involved where a component does not behave as expected.
\end{itemize}

\subsubsection{Suggested Readings and Resources}

In the following, we list a set of resources (data and tools) that might be useful in building such a system. 

\noindent
Software platforms:
\begin{itemize}
\item 
Macaw: A conversational information seeking platform implemented in Python which supports multiple interfaces and modalities~\cite{Zamani:2019}. 
\item 
TIRA Integrated Research Architecture \cite{Potthast:2019} (a modularized platform for shared tasks).
\end{itemize}

\noindent
Scientific IR tools:
\begin{itemize}
\item 
ArXivDigest: A personalized scientific literature recommendation framework based on arXiv articles.\footnote{\url{https://github.com/iai-group/arxivdigest}}
\item 
GrapAL: Querying Semantic Scholar's literature graph \cite{Betts:2019} (web-based tool for exploring scientific literature, e.g., finding experts on a given topic).\footnote{\url{https://allenai.github.io/grapal-website/}}
\end{itemize}

\noindent
Open-source scholarly conversational agents:
\begin{itemize}
\item 
UKP-ATHENA: A scientific conversational agent \cite{Mesgar:2019} (early prototype for assisting ACL{*} conference attendees and answering basic ACL Anthology queries).\footnote{\url{http://athena.ukp.informatik.tu-darmstadt.de:5002/}}
\end{itemize}

Data collections suitable to be incorporated in the Scholarly Conversational Assistant: 
\begin{itemize}
\item 
Open Research Knowledge Graph\footnote{\url{http://orkg.org}} (ORKG)~\cite{Jaradeh:2019:ORK:3360901.3364435}: Semantic annotations of publications 
\item 
Semantic Scholar: Articles in a broad range of fields
\item 
ACM DL: A subset of computer science articles
\item 
dblp: A clean list of computer science articles
\item 
ACL Anthology: A public collection of ACL{*} articles
\item 
Open Review: A small subset of conference articles with public reviews
\item 
Other sources: Google Scholar, Citeseer, Arnetminer, Conference apps (e.g., Whova)
\end{itemize}

\noindent
Other related work:
\begin{itemize}
\item 
\cite{Gentile:2015:CLA}: Recupero: Conference Live: Accessible and Sociable Conference Semantic Data
\item 
\cite{Dalton:2018:VGC}: Vote Goat: Conversational Movie Recommendation
\item 
\cite{wan2019aminer}: Aminer: Search and mining of academic social networks (researcher-centric IR)
\end{itemize}

\newpage
\makeatletter
\addtocontents{toc}{\let\string\@secpagenumber\relax}%
\makeatother

\section{Recommended Reading List}
\label{reading-list}

These publications were recommended by the seminar participants via the pre-seminar survey. Please also refer to the reading list available in individual reports of working groups.

\begin{itemize}
\item
Saleema Amershi, Dan Weld, Mihaela Vorvoreanu, Adam Fourney, Besmira Nushi, Penny Collisson, Jina Suh, Shamsi Iqbal, Paul N. Bennett, Kori Inkpen, Jaime Teevan, Ruth Kikin-Gil, and Eric Horvitz. Guidelines for Human-AI Interaction, \textsl{CHI 2019}. \url{https://doi.org/10.1145/3290605.3300233}
\item
Aliannejadi, Mohammad, Hamed Zamani, Fabio Crestani, and W. Bruce Croft. Asking clarifying questions in open-domain information-seeking conversations. \textsl{SIGIR 2019}. \url{https://doi.org/10.1145/3331184.3331265}
\item
Belkin, Nicholas J., Colleen Cool, Adelheit Stein, and Ulrich Thiel.  Cases, scripts, and information-seeking strategies: On the design of interactive information retrieval systems. \textsl{Expert systems with applications}, 9 (3), 1995. \url{https://doi.org/10.1016/0957-4174(95)00011-W}
\item
Timothy Bickmore, Justine Cassell. Social dialogue with embodied conversational agents. Advances in natural multimodal dialogue systems, 2005. \url{https://doi.org/10.1007/1-4020-3933-6_2}
\item
Daniel Braun, Adrian Hernandez-Mendez, Florian Matthes, Manfred Langen. Evaluating natural language understanding services for conversational question answering systems. \textsl{SIGdial 2017}. \url{https://doi.org/10.18653/v1/W17-5522}
\item
Andrew Breen, et al. Voice in the User Interface. Interactive Displays: Natural Human-Interface Technologies, 2014, \url{https://doi.org/10.1002/9781118706237.ch3.}
\item
Brennan, Susan E., and Eric A. Hulteen. Interaction and feedback in a spoken language system: A theoretical framework. \textsl{Knowledge-based systems}, 8 (2-3), 1995. \url{https://doi.org/10.1016/0950-7051(95)98376-H}
\item
Harry Bunt. Conversational principles in question-answer dialogues. \textsl{Zur Theorie der Frage}, pages 119-141.
\item
Justine Cassell. Embodied conversational agents. \textsl{AI Magazine}, 22(4), 2001. \url{https://doi.org/10.1609/aimag.v22i4.1593}
\item
Justine Cassell, Joseph Sullivan, Elizabeth Churchill, Scott Prevost. Embodied conversational agents. MIT Press, 2000.
\item
Eunsol Choi, He He, Mohit Iyyer, Mark Yatskar, Wen-tau Yih, Yejin Choi, Percy Liang, Luke Zettlemoyer. QuAC: Question Answering in Context. \textsl{EMNLP 2018}. \url{https://dx.doi.org/10.18653/v1/D18-1241}
\item
Christakopoulou, Konstantina, Filip Radlinski, and Katja Hofmann. Towards conversational recommender systems. \textsl{SIGKDD 2016}. \url{https://doi.org/10.1145/2939672.2939746}
\item
Leigh Clark, Phillip Doyle, Diego Garaialde, Emer Gilmartin, Stephan Schlögl, Jens Edlund, Matthew Aylett, João Cabral, Cosmin Munteanu, Benjamin Cowan. The State of Speech in HCI: Trends, Themes and Challenges. \textsl{Interacting with Computers}, 31 (4), 2019. \url{https://doi.org/10.1093/iwc/iwz016}
\item
Mark Core and James Allen. Coding dialogs with the DAMSL annotation scheme. \textsl{AAAI Fall Symposium on Communicative Action in Humans and Machines}, 1997.
\item
Paul Grice. Studies in the Way of Words. Harvard University Press, 1989.
\item
Haider, Jutta, and Olof Sundin. Invisible Search and Online Search Engines: The ubiquity of search in everyday life. Routledge, 2019.
\item
Ben Hixon, Peter Clark, Hannaneh Hajishirzi. Learning knowledge graphs for question answering through conversational dialog. \textsl{NAACL 2015}. \url{http:s//dx.doi.org/10.3115/v1/N15-1086}
\item
Mohit Iyyer, Wen-tau Yih, Ming-Wei Chang. Search-based neural structured learning for sequential question answering. \textsl{ACL 2017}. \url{https://dx.doi.org/10.18653/v1/P17-1167}
\item
Diane Kelly and Jimmy Lin. Overview of the TREC 2006 ciQA task. \textsl{SIGIR Forum} 41(1), 2007. \url{https://doi.org/10.1145/1273221.1273231}
\item
Liu, Bei, Jianlong Fu, Makoto P. Kato, and Masatoshi Yoshikawa. Beyond narrative description: Generating poetry from images by multi-adversarial training. \textsl{ACM Multimedia}, 2018. \url{https://doi.org/10.1145/3240508.3240587}
\item
Dominic W. Massaro, Michael M. Cohen, Sharon Daniel, Ronald A Cole. Developing and evaluating conversational agents. \textsl{Human performance and ergonomics}, 1999. \url{https://doi.org/10.1016/B978-012322735-5/50008-7}
\item
McTear, Michael F. Spoken dialogue technology: enabling the conversational user interface.\textsl{ ACM Computing Surveys}, 34 (1), 2002. \url{https://doi.org/10.1145/505282.505285}
\item
Oddy, Robert N. Information retrieval through man-machine dialogue. Journal of documentation, 33 (1), 1977. \url{https://doi.org/10.1108/eb026631}
\item
Filip Radlinski, Nick Craswell. A theoretical framework for conversational search. CHIIR 2017. \url{https://doi.org/10.1145/3020165.3020183}
\item
Siva Reddy, Danqi Chen, Christopher D. Manning. CoQA: A conversational question answering challenge. \textsl{TACL, 7, 2019}. \url{https://doi.org/10.1162/tacl_a_00266 }
\item
Ren, Gary, Xiaochuan Ni, Manish Malik, and Qifa Ke. Conversational query understanding using sequence to sequence modeling. \textsl{The Web Conference}, 2018. \url{https://doi.org/10.1145/3178876.3186083}
\item
Zs\'{o}fia Ruttkay, Catherine Pelachaud. From brows to trust: Evaluating embodied conversational agents. \textsl{Springer Science \& Business Media}, 2004. \url{https://doi.org/10.1007/1-4020-2730-3}
\item
Shum, Heung-Yeung, Xiao-dong He, and Di Li. From Eliza to XiaoIce: challenges and opportunities with social chatbots. Frontiers of Information Technology \& Electronic Engineering, 19 (1), 2018. \url{https://doi.org/10.1631/FITEE.1700826}
\item
Adelheit Stein, Elisabeth Maier. Structuring Collaborative Information-Seeking Dialogues. \textsl{Knowledge-Based Systems}, 8(2-3), 1995. \url{https://doi.org/10.1016/0950-7051(95)98370-L}
\item
Oriol Vinyals, Quoc Le. A neural conversational model. ICML Deep Learning Workshop, 2015. \url{https://arxiv.org/abs/1506.05869}
\item
Marylyn Walker, Dianne Litman, Candace Kamm, Alicia Abella. PARADISE: A framework for evaluating spoken dialogue agents. \textsl{ACL 1997}. \url{https://dx.doi.org/10.3115/976909.979652}
\item
Weston, J., Bordes, A., Chopra, S., Rush, A.M., van Merri\"{e}nboer, B., Joulin, A. and Mikolov, T. Towards ai-complete question answering: A set of prerequisite toy tasks. \url{https://arxiv.org/abs/1502.05698}
\item
Wu, Wei, and Rui Yan. Deep Chit-Chat: Deep Learning for Chit-Chat. \textsl{SIGIR 2019}. \url{https://doi.org/10.1145/3331184.3331388}
\item
Zhang, Yongfeng, Xu Chen, Qingyao Ai, Liu Yang, and W. Bruce Croft. Towards conversational search and recommendation: System ask, user respond. \textsl{CIKM 2018}. \url{https://doi.org/10.1145/3269206.3271776}
\item
Kangyan Zhou, Shrimai Prabhumoye, and Alan W Black. A dataset for document grounded conversations. \textsl{EMNLP 2018}. \url{https://dx.doi.org/10.18653/v1/D18-1076}
\item
Zhou, Li, Jianfeng Gao, Di Li, and Heung-Yeung Shum. The design and implementation of XiaoIce, an empathetic social chatbot. \textsl{Computational Linguistics}, 2020. \url{https://doi.org/10.1162/coli_a_00368}
\end{itemize}

\section{Acknowledgements}

The seminar organisers would like to thank all participants and speakers of invited talks for their active contributions. We also thank the staff of Schloss Dagstuhl for providing a great venue for a successful seminar. The organisers were in part supported by JSPS KAKENHI Grant Number 19H04418. Any opinions, findings, and conclusions described here are the authors and do not necessarily reflect those of the sponsors.

\begin{participants}
\participant Khalid Al-Khatib \\Bauhaus University Weimar, DE
\participant Avishek Anand \\Leibniz Universit\"at Hannover, DE
\participant Elisabeth Andr\'e \\University of Augsburg, DE
\participant Jaime Arguello \\University of North Carolina at Chapel Hill, US
\participant Leif Azzopardi \\University of Strathclyde -- Glasgow, GB
\participant Krisztian Balog \\University of Stavanger, NO
\participant Nicholas J. Belkin \\Rutgers University -- New Brunswick, US
\participant Robert Capra \\University of North Carolina at Chapel Hill, US
\participant Lawrence Cavedon \\RMIT University -- Melbourne, AU
\participant Leigh Clark \\Swansea University, UK
\participant Phil Cohen \\Monash University -- Clayton, AU
\participant Ido Dagan \\Bar-Ilan University -- Ramat Gan, IL
\participant Arjen P. de Vries \\Radboud University Nijmegen, NL
\participant Ondrej Dusek \\Charles University -- Prague, CZ
\participant Jens Edlund \\KTH Royal Institute of Technology -- Stockholm, SE
\participant Lucie Flekova \\Amazon R\&D -- Aachen, DE
\participant Bernd Fröhlich \\Bauhaus University Weimar, DE
\participant Norbert Fuhr \\University of Duisburg--Essen, DE
\participant Ujwal Gadiraju \\Leibniz Universit\"at Hannover, DE
\participant Matthias Hagen \\Martin Luther University Halle--Wittenberg, DE
\participant Claudia Hauff \\TU Delft, NL
\participant Gerhard Heyer \\University of Leipzig, DE
\participant Hideo Joho \\University of Tsukuba -- Ibaraki, JP
\participant Rosie Jones \\Spotify -- Boston, US
\participant Ronald M. Kaplan \\Stanford University, US
\participant Mounia Lalmas \\Spotify -- London, GB
\participant Jurek Leonhardt \\Leibniz Universit\"at Hannover, DE
\participant David Maxwell \\University of Glasgow, GB
\participant Sharon Oviatt \\Monash University -- Clayton, AU
\participant Martin Potthast \\University of Leipzig, DE
\participant Filip Radlinski \\Google UK -- London, GB
\participant Rishiraj Saha Roy \\MPI for Computer Science -- Saarbrücken, DE
\participant Mark Sanderson \\RMIT University -- Melbourne, AU
\participant Ruihua Song \\Microsoft XiaoIce -- Beijing, CN
\participant Laure Soulier \\UPMC -- Paris, FR
\participant Benno Stein \\Bauhaus University Weimar, DE
\participant Markus Strohmaier \\RWTH Aachen University, DE
\participant Idan Szpektor \\Google Israel -- Tel Aviv, IL
\participant Jaime Teevan \\Microsoft Corporation -- Redmond, US
\participant Johanne Trippas \\RMIT University -- Melbourne, AU
\participant Svitlana Vakulenko \\Vienna University of Economics and Business, AT
\participant Henning Wachsmuth \\University of Paderborn, DE
\participant Emine Yilmaz \\University College London, UK
\participant Hamed Zamani \\Microsoft Corporation, US
\end{participants}

\bigskip
\begin{center}
\includegraphics[width=0.8\textwidth]{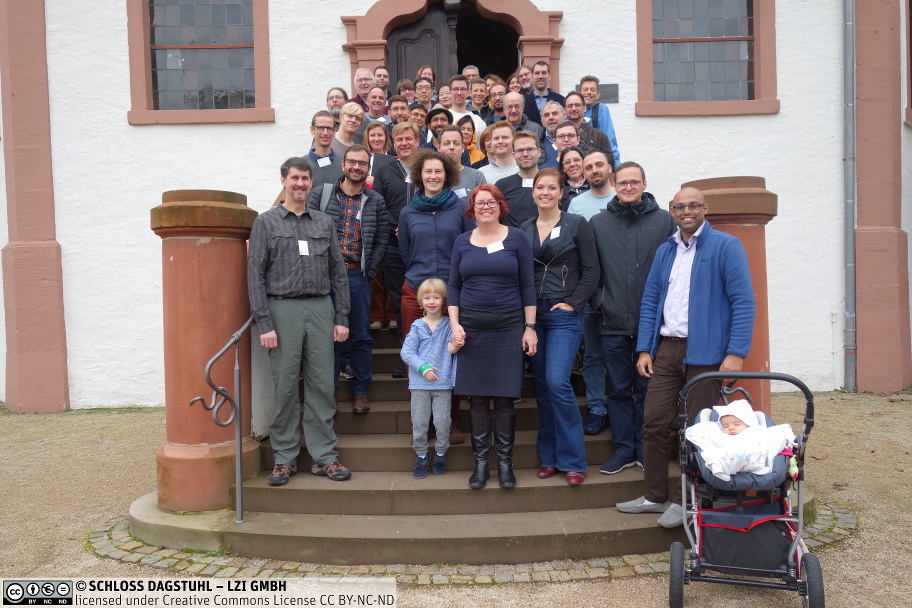}
\end{center}


\end{document}